\documentclass[fleqn,usenatbib]{mnras}

\usepackage{newtxtext,newtxmath}

\usepackage[T1]{fontenc}
\usepackage{ae,aecompl}
\usepackage{array}
\usepackage{color}
\usepackage{textcase}

\usepackage{gensymb}
\usepackage{graphicx}	
\usepackage{amsmath}	
\usepackage{footmisc}

\title{Predictions for the FAST telescope's CRAFTS Extra-Galactic HI Survey}

\author[K. Zhang et al.]{Kai Zhang,$^{1,2}$\thanks{E-mail: zk3kw2n@nao.cas.cn}, Jingwen Wu$^{1,2}$\thanks{E-mail: jingwen@nao.cas.cn}, Di Li$^{1,2,3}$\thanks{E-mail:dili@nao.cas.cn}, Chao-Wei Tsai$^{1,2}$, Lister Staveley-Smith$^{4,5}$, \newauthor Jing Wang$^{6}$, Jian Fu$^{7}$, Travis McIntyre$^{1}$, Mao Yuan$^{1,2}$ and FAST collaboration
\\
\\
$^{1}$National Astronomical Observatories, Chinese Academy of Sciences, Datun Road, Chaoyang District, Beijing 100101, China;\\
$^{2}$University of Chinese Academy of Sciences, Beijing 100049, China;\\
$^{3}$NAOC-UKZN Computational Astrophysics Centre, University of KwaZulu-Natal, Durban 4000, South Africa;\\
$^{4}$International Centre for Radio Astronomy Research (ICRAR), The University of Western Australia, 35 Stirling Hwy, Crawley, WA 6009, Australia;\\
$^{5}$ARC Centre of Excellence for All Sky Astrophysics in 3 Dimensions (ASTRO 3D), Australia;\\
$^{6}$Kavli Institute for Astronomy and Astrophysics, Peking University, Beijing 100871, China;\\
$^{7}$Key Laboratory for Research in Galaxies and Cosmology, Shanghai Astronomical Observatory, CAS, 80 Nandan Road., Shanghai, 200030, China
}

\date{Accepted XXX. Received YYY; in original form ZZZ}

\pubyear{2015}

\begin{document}
\label{firstpage}
\pagerange{\pageref{firstpage}--\pageref{lastpage}}
\maketitle

\begin{abstract}

The Five-hundred-meter Aperture Spherical radio Telescope (FAST) has started the Commensal Radio Astronomy FasT Survey (CRAFTS). In this paper, we use the technical parameters of FAST derived from commissioning observations to simulate the completeness function for extragalactic HI survey of CRAFTS, HI galaxies from two kinds of mock catalogues are selected. One is generated by Monte-Carlo simulation based on the interpolated mass-velocity width function of the ALFALFA $100\%$ (a.k.a. $\alpha$.100) catalogue. The other is constructed by semi-analytical N-body simulation based on the $\Lambda$CDM model. Our results suggest that a two-pass extragalactic HI survey will be able to detect nearly $4.8\times10^{5}$ galaxies, from which the “faint end” slope of the HI Mass Function (HIMF) can be recovered to $\mathrm{10^{7}\,M_{\odot}}$ and the "knee mass" of the HIMF can be measured to a redshift of 0.1. Considering the radio frequency interference status and sensitivity limitation, CRAFTS will be efficient in detecting HI galaxies at redshifts below 0.1, which implies a tremendous potential in exploring the galaxy interactions in different environments and the spatial distribution of HI galaxies in the local universe.

\end{abstract}


%%%%%%%%%%%%%%%%%%%%%%%%%%%%%%%%%%%%%%%%%%%%%%%%%%

%%%%%%%%%%%%%%%%% BODY OF PAPER %%%%%%%%%%%%%%%%%%

\section{INTRODUCTION}

%=======================

\begin{table*}%[t]
\begin{center}
\begin{tabular}{ l c c}
\hline
Survey strategy & \multicolumn{2}{c}{2-pass drift scan}\\
Scan spacing & \multicolumn{2}{c}{$\mathrm{21.9' \, in\, declination }$}\\ 
Sky coverage  &  \multicolumn{2}{c}{$\mathrm{\sim 20000\,deg^2 }$}\\  
Declination range & \multicolumn{2}{c}{$\mathrm{-14 \degree \,to\,+66\degree }$} \\ 
Number of beams & \multicolumn{2}{c}{19} \\
Polarizations per beam & \multicolumn{2}{c}{$\mathrm{full\,stokes}$} \\
Frequency range & \multicolumn{2}{c}{$\mathrm{1050-1450\,MHz}$} \\
Channel width & \multicolumn{2}{c}{$\mathrm{7.6\,kHz\,(1.6 \,km\,s^{-1}\,at\,1420.4058\,MHz )}$}   \\
System temperature & \multicolumn{2}{c}{$\mathrm{18-26\,K}$} \\ \hline
Zenith Angle (ZA) & $\mathrm{0-26.4\degree }$  & $\mathrm{26.4 \degree-40\degree }$\\ 
Effective illuminated aperture size & $\mathrm{300\,m}$  & $\mathrm{250-300\,m}$\\ 
Gain (center beam) & $\mathrm{16.5\,K\,/\,Jy}$  & $\mathrm{11.0-16.5\,K\,/\,Jy}$\\
Beamsize (FWHM) at 1420.4058MHz & $\mathrm{2.95'}$ & $\mathrm{2.95'-3.60'}$ \\ %\hline
\hline
\end{tabular}
\caption{The technical parameters of CRAFTS extragalactic HI survey derived from commissioning observations. The two columns at the bottom reflect the change in sensitivity for the telescope at low and high zenith angles (ZAs). At ZAs above $26.4\degree$, the reflecting surface is only partially illuminated. For detail descriptions, please see \citealt{Zhang2019}. }
\label{tab:CRAFTS}
\end{center}
\end{table*}

%=======================

The current standard $\Lambda$ cold dark matter ($\mathrm{\Lambda CDM}$) cosmological model has successfully simulated the formation history of galaxies and the cosmic structure on large scales, but there are still several controversial discrepancies between simulations and observations at small scales \citep{Weinberg2015}. One puzzle is that the observed low-mass galaxies (satellites) are much rarer than the number of subhalos predicted by the model, which is usually referred to as the "missing satellites" problem \citep{Klypin1999,Moore1999}. The "missing satellites" problem can also be reflected in the discrepancy of the derived dark matter halo mass function between simulations and observations. In the $\mathrm{\Lambda CDM}$ model \citep{Press1974}, dark matter halo mass function can be well fitted by Schechter function, with a low-mass end or "faint end" of  $\alpha \sim -$1.9. However, from existing galaxy surveys, the "faint end" of the optical luminosity function (LM) and the HI Mass Function (HIMF) are much flatter, typically between $-1.0$ and $-1.4$ (e.g. \citealt{Blanton2005,Zwaan2005,Montero2009,Hill2010,Martin2010,Jones2018}). The overpredicted number of satellites might be alleviated by adding baryonic processes like photoionization and stellar feedback (e.g. \citealt{Koposov2009,Maccio2010}) and by considering tidal interactions with the Galactic disc (e.g., \citealt{Garrison2017,Nadler2018,Kelley2019}) in simulations. The census of dwarf galaxies has also improved due to the enhancement of survey sensitivity and sky coverage (e.g. \citealt{Tollerud2008,Drlica2019}). However, this mismatch has not been fully eliminated. The HI observations provide an alternate approach to identify dwarf galaxies because of their higher neutral gas fraction \citep{Schombert2001,Geha2006,Haynes2019}.

Despite the "missing satellites" problem, recent observations of the most luminous satellites around the Milky Way indicate that the dark matter subhaloes predicted from the $\mathrm{\Lambda CDM}$ simulation are more massive than the results inferred from stellar kinematics in the satellites, which is referred to as the "Too Big To Fail" (TBTF) problem \citep{Boylan-Kolchin2011,Boylan-Kolchin2012}. This problem is also discovered in the M31 system \citep{Tollerud2014} and field galaxies \citep{Ferrero2012,Papastergis2015}, which implies that this may not be caused by environmental interactions. The dark matter halo mass function can be predicted by the rotational velocity function of galaxies and the rotational velocity of a galaxy can be inferred from its velocity width of HI profile, as HI can extend farther in the radial direction. Larger samples are required to further investigate the origin of this phenomenon.

The large blind HI survey is the most efficient way in studying the statistical properties of HI galaxies like the HIMF, the velocity width function (WF) and the two-point correlation function (2PCF). The HIPASS (HI Parkes All Sky Survey; \citealt{Barnes2001,Meyer2004}) and the ALFALFA (Arecibo Legacy Fast ALFA; \citealt{Giovanelli2005,Haynes2011,Giovanelli2015,Haynes2018}) survey are the two largest HI surveys to date, which have successfully provided the HI distribution properties in a fair cosmological volume. However, new questions have been raised due to the limited sensitivity and/or sky coverage. For example, from the HIPASS and $\alpha.40$ catalogues, no obvious clustering dependence of galaxies on HI mass has been identified by measuring the projected 2PCF \citep{Basilakos2007,Meyer2007,Martin2012,Papastergis2013}. However, \cite{Guo2017} found a strong positive correlation between HI mass and clustering properties through the effective volume-limited projected 2PCF measurements of the ALFALFA $70\%$ (a.k.a. $\alpha.70$) catalogue. The clustering measurements in the latter work are biased by the superclusters at HI masses below $\mathrm{10^9 \, M_{\odot}}$, which demands a larger survey volume and sample size to achieve more robust measurements. In addition, as noted in \cite{Jones2018}, the "faint end" slope of the HIMF derived from the known HI-deficient Virgo cluster is flatter than that from the galaxies in its immediate surroundings. They hypothesized that the Virgo cluster might be in a gas-rich environment and the surrounded gas-rich filaments might feed the growth of the cluster, which needs deeper and wider observations to detect the low-mass galaxies in that region.

The Five-hundred-meter Aperture Spherical radio Telescope (FAST; \citealt{Nan2011, Li2016}) is now the largest filled-aperture single dish \citep{Jiang2020} and aims to carry out large-scale HI and pulsar surveys \citep{Li2019}. FAST plans to achieve high-quality HI imaging and pulsar searching simultaneously during observations by injecting a high-frequency electronic noise signal for calibration (CAL; \citealt{Li2018}). The survey plan is designated as the Commensal Radio Astronomy FAST Survey (CRAFTS). Benefitting from this innovative observation mode,  CRAFTS will be able to conduct extragalactic HI, galactic HI imaging, pulsar search, and fast radio burst (FRB) search surveys simultaneously. The large collecting area and great sensitivity of FAST will allow a superior detectability on the HI gas in the extra-galactic systems \citep{Duffy2008,Zhang2019} and provide a high quality data set for investigating the aforementioned issues. In the following sections unless specified, the CRAFTS refers to the two-pass CRAFTS extragalactic HI galaxy survey.

This paper is organized as follows: Section~\ref{sec:sec2} describes the technical parameters of CRAFTS extragalactic HI survey and HI mass detection limit. In Section~\ref{sec:sec3} and ~\ref{sec:sec4}, we estimate the number of HI galaxies CRAFTS may detect from Monte-Carlo and semi-analytic simulation, respectively. Sections~\ref{sec:sec6} and ~\ref{sec:sec5} discuss the recovered HIMF and source confusion rate from mock catalogues. Section~\ref{sec:sec9} illustrates our improved normalization method of the HIMF, which is less susceptible to the selective effects of the survey. In Section~\ref{sec:sec7}, we discuss the influence of source confusion and radio frequency interference (RFI) on the survey. We will summarize our predicted results of CRAFTS in Section~\ref{sec:sec8}. 

In this paper, we adopt a cosmology of $ H_{0}$ (Hubble constant) = 70 $\mathrm{km\, s^{-1} \, Mpc^{-1}}$, $\Omega_{\rm M}$ (density parameter of matter) = 0.272 and $\Omega_{\Lambda}$ (density parameter of dark energy) = 0.728 \citep{Komatsu2011}. The calculations of cosmological distances in this paper are mainly referred to \cite{Hogg1999} and \cite{Meyer2017}.

\section{THE SIMULATED CRAFTS EXTRA-GALACTIC HI SURVEY }
\label{sec:sec2}

\subsection{Technical parameters of CRAFTS}

As illustrated in \cite{Giovanelli2005}, to achieve more detections, increasing the survey solid angle is more efficient than increasing the integration time. CRAFTS aims to complete a relatively shallow but wide survey within a declination (DEC) range between $-14 \degree$ and $66\degree$, covering over $20000\,\mathrm{deg^{2}}$ of sky regions and reaching a redshift of 0.35. As discussed in \cite{Li2018} and \citet[hereinafter Zh19]{Zhang2019}, the aperture of FAST can be fully illuminated at zenith angles (ZAs) up to $26.4 \degree $ and is only partially illuminated at ZAs up to $40 \degree$, so there is gain loss and an increase in beam size when ZA is above $26.4 \degree$. The system temperature of FAST will increase with ZA because of the radiation due to the surrounding mountain peaks entering the nearer sidelobes. To minimize the influence of gain fluctuation and RFI to the survey, CRAFTS plans to conduct a full two-pass drift-scan survey by the FAST L-band Array of 19-beam (FLAN) receiver, which will be rotated by $23.4 \degree$ to achieve a super-Nyquist sampling while drifting \citep{Li2018}. In the now formally approved CRAFTS observing program, there is only time allowed for one pass. We aim to supplement the observation with open time proposals, which intends to be scheduled for high valued sky area (e.g. with clusters) about 6 months apart from the survey epoch.

The basic parameters of the CRAFTS extragalactic HI survey are summarized in Table~\ref{tab:CRAFTS}. The technical parameters used in this paper, including gain, system temperature and beam size, were obtained from commissioning observations and had been summarized in \citet{Zhang2019}. Note that the effective integration time calculated in this paper is similar to Equation (6) in Zh19, but we used the noise-equivalent beam area (NEBA) instead of the beam area as in Zh19. Since when combining data from multiple observations, the weighting factor should be the square of the beam response, in order to maximize the signal-to-noise ratio (SNR) \citep{Staveley1997}. Therefore, the noise pattern will differ from the beam pattern and the NEBA will be a half of the beam area. As a result, the effective integration time in Zh19 is overpredicted by a factor of 2 from this point of view, and the sensitivity is over-predicted by a factor of $\sqrt{2}$, which leads the predicted average channel rms of two-pass CRAFTS to increase from 1.18 to 1.67 $\mathrm{mJy}$. For comparison, after converting to the same velocity resolution of HIPASS and ALFALFA, the effective rms noise of CRAFTS is expected to be nearly 0.47 $\mathrm{mJy}$ (the sensitivity of HIPASS is 13 $\mathrm{mJy}$ with 26 $\mathrm{km\,s^{-1}}$ resolution; \citealt{Barnes2001}) and 0.67 $\mathrm{mJy}$ (the sensitivity of ALFALFA is 2.0 $\mathrm{mJy}$ with 10 $\mathrm{km\,s^{-1}}$ resolution; \citealt{Haynes2018}), respectively.

\subsection{Flux limit for HI detection}

For FAST, it's reasonable to assume that most of the observed HI galaxies are point sources \citep{Duffy2008}. If we ignore the peculiar motion of the Earth and HI sources, the frequency of an HI line we observe is $\nu_{\rm obs}=\nu_{\rm HI}/(1+z)$, where $\nu_{\rm HI} =1420.4058 \,\rm MHz$ is the rest-frame HI frequency, and $ z$ is the redshift of the source. Using the non-relativistic assumption, the line-of-sight velocity can be approximated by $cz$, where $c$ is the speed of light.

The velocity width of an HI line profile can be measured at $\rm 50\%$ level of each of two peaks or the full width of half maximum (FWHM) of one peak, which is termed as $\mathrm{W_{50}}$. The velocity width observed by the telescope is measured in the observed frame while the intrinsic velocity width of an HI source refers to the velocity width in the source's rest frame, which is termed as $W_{\mathrm{obs}}$ and $ W_{ \mathrm{intri} }$, respectively. Then the relation between $W_{\mathrm{obs}}$ and $ W_{ \mathrm{intri} }$ can be denoted by \citep{Peacock1999, Abdalla2005, Meyer2017}
\begin{eqnarray}
\label{eq:W_obs}
W_{\mathrm{obs}}=(1+z)W_{\mathrm{intri}}, 
\end{eqnarray}
where $W_{\mathrm{obs}}$ and $W_{\mathrm{intri}}$ are in $\mathrm{km \, s^{-1}}$.

The corresponding velocity width $\Delta V_{\mathrm {ch}}$ in observed frame within a channel width of $\Delta f_{\mathrm{ch}}$ can be estimated by \citep{Meyer2017}
\begin{eqnarray}
\label{eq:V_ch}
\Delta V_{\mathrm {ch}} \simeq \frac{c(1+{z})^{2}}{\nu_{\mathrm {HI} }}{\Delta f_{\mathrm {ch}}}.
\end{eqnarray}We assume the peculiar motion speed and the velocity width of the source are much less than $c$. For simplicity, when calculating the velocity integrated flux of an HI line in terms of observed frame velocity using mock catalogues, we assume a top-hat line profile with a peak flux $S_{\mathrm{\,\nu}}^{\mathrm{peak}}$, so that $S^{\mathrm{obs}}_{\mathrm{int}}=S_{\mathrm{\,\nu}}^{\mathrm{peak}}W_{\mathrm{obs}}$.

The SNR can be calculated by \citep{Haynes2011}
\begin{eqnarray}
\label{eq:SNR}
\mathrm{SNR} = S_{\mathrm{\,\nu}}^{\mathrm{peak}}\left(\frac{S_{\mathrm{rms}}}{f_{\mathrm{smo}}^{1/2}}\right)^{-1},
\end{eqnarray}where $ S_{\mathrm{rms}}$ is the rms noise per channel, $ f_{\mathrm{smo}}$ is the number of independent channels that the source signal can be smoothed over. As illustrated in \citet{Giovanelli2005}, the SNR is best rendered when the noise is measured after smoothing a signal to a spectral resolution of about half of its linewidth, thus $f_{\mathrm{smo}}$\footnote{After conformed by M. G. Jones privately, there might be a typo in the interpretation of $f_{\mathrm{smo}}$ in \citet{Jones2015}.} can be given by \citep{Jones2015} 
\begin{eqnarray}
\label{eq:f_smo}
f_{\mathrm{smo}} = \frac{1}{2*\Delta V_{\mathrm{ch}}} \left\{
\begin{array}{ll}
W_{\mathrm{obs}} & \mathrm{if} \; W_{\mathrm{obs}} \leq W_{\rm c} \\
W_{\rm c} & \mathrm{if} \; W_{\mathrm{obs}} > W_{\rm c},
\end{array}
\right.
\end{eqnarray}where $\log{W_{\rm c}/\mathrm{km\,s^{-1}}}=2.5$ is the transition caused by the very broad spectral "standing waves" resulting from the reflections in the telescope focal structure (e.g. \citealt{Briggs1997,Haynes2011}). Thus the velocity integrated flux limit for detecting an HI galaxy in terms of observed frame velocity can be estimated by
\begin{eqnarray}
\label{eq:Sint_lim}
S^{\mathrm{obs}}_{\mathrm{int,lim}}=\rm SNR_{\mathrm{lim}}*S_{\mathrm{rms}}W_{\mathrm{obs}}/f_{\mathrm{smo}}^{\mathrm{1/2}},
\end{eqnarray}
where $\rm SNR_{\mathrm{lim}}$ is the limiting $\rm SNR$ for detections.

%=================
\begin{figure}
	\includegraphics[scale=0.40]{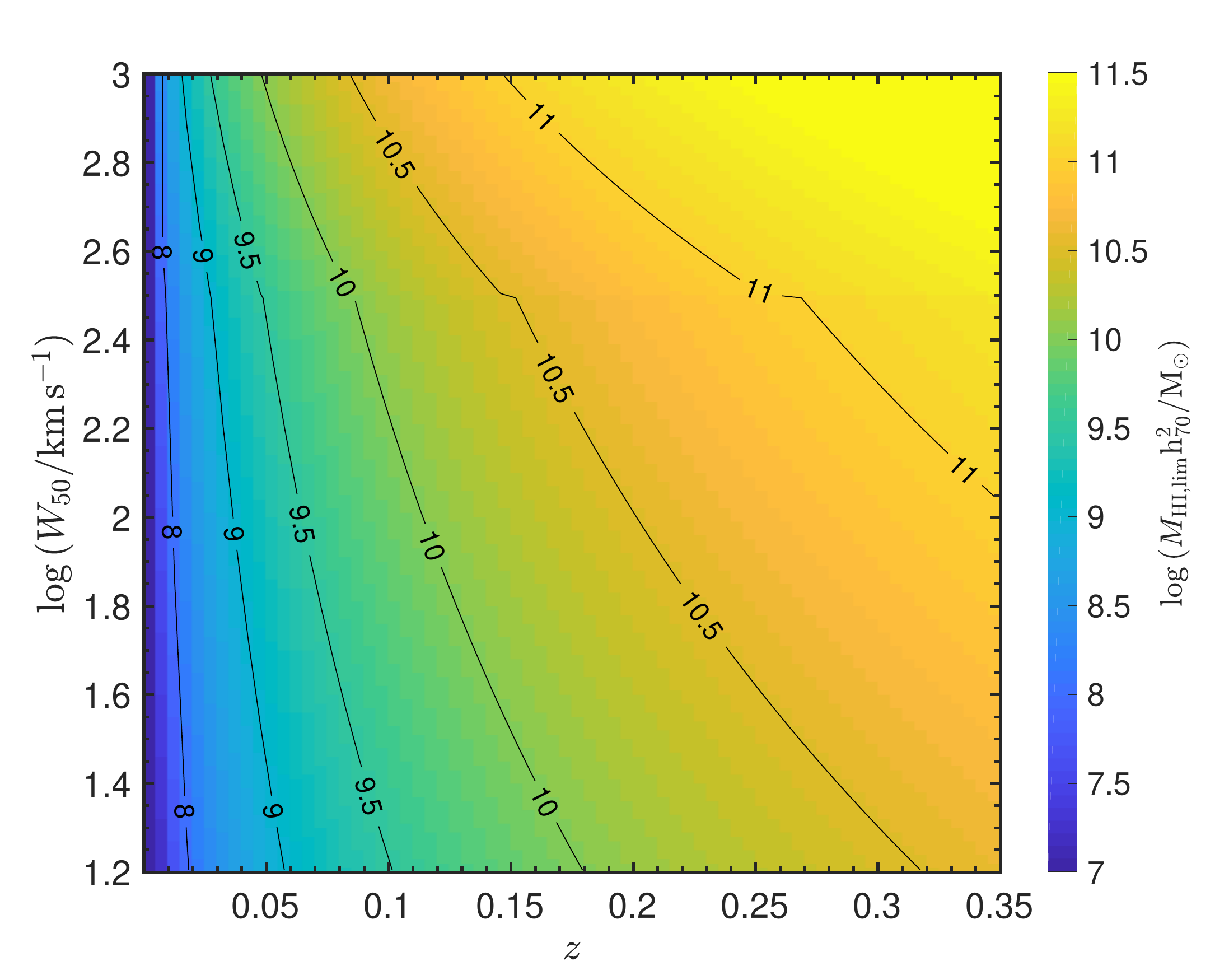}
    \caption{The HI mass limit of two-pass CRAFTS survey in the redshift and intrinsic velocity width plane at a ZA of $0\degree$. The solid lines represent the HI mass limit contours.}
    \label{fig:M_lim1}
\end{figure}
%+================

By combining equations~\ref{eq:W_obs},~\ref{eq:V_ch},~\ref{eq:SNR},~\ref{eq:f_smo}, and~\ref{eq:Sint_lim}, one can obtain the velocity integrated flux limit in observed frame for detecting an HI galaxy at a redshift of $z$ and with an intrinsic velocity width of $W_{\mathrm{intri}}$, which can be denoted by 
\begin{eqnarray}
\label{eq:S_intlim}
&&\log{  \left(\frac{  S^{\mathrm{obs}}_{\mathrm{int,lim}} }  {\mathrm{Jy\, km\, s^{-1}}}\right) }= \log{\rm SNR_{lim}}+\log{  \left(\frac{S_{\mathrm{rms}}}{\mathrm{Jy}}\right) } + \nonumber\\
&& \frac{1}{2}\left[\log{\left(\frac{c}{\mathrm{km\,s^{-1}}} \right)}+ \log{\left(\frac{\Delta f_{\mathrm{ch}}}{\mathrm{kHz}}\right) } -         \log{\left(\frac{\nu_{\mathrm{HI}}}{\mathrm{kHz}} \right)}\right]+ \nonumber\\
&& 
\left\{
\begin{array}{ll}
\frac{1}{2}\log{\left(\frac{W_{\mathrm{intri}}}{\mathrm{km \,s^{-1}}}\right)}+\frac{3}{2}\log{(1+z)} & \mathrm{if} \; W_{\mathrm{obs}} \leq W_{\rm c} \\
\log{\left(\frac{W_{\mathrm{intri}}}{\mathrm{km \, s^{-1}}}\right)}+2\log{(1+z)}-1.25 & \mathrm{if} \; W_{\mathrm{obs}} > W_{\rm c}.
\end{array}{}
\right.
\end{eqnarray}

The completeness of a survey can be defined as the fraction of cosmic sources from the underlying distributions that are detected by the survey. The specific form of the completeness function will differ from different smoothing and source extraction methods. For ALFALFA survey, the completeness function is defined by the relation between the velocity width and  integrated flux density of HI galaxies \citep{Haynes2011}. As described in \cite{Rosenberg2002} and \cite{Haynes2011}, using a $50\%$ completeness limit and a full completeness limit as the selection criterion obtains approximately similar statistical results. We estimate the selection function of CRAFTS by referring to the ALFALFA $50\%$ completeness function. By replacing the sensitivity per channel of ALFALFA $40\%$ (3.4 $\mathrm{mJy}$ per 24.4 $\mathrm{kHz}$ \footnotemark; \citealt{Haynes2011,Jones2015}) \footnotetext{Noted that the average rms of ALFALFA has been revised in ALFALFA $100\%$  \citep{Haynes2018}, which is 2.8 $\rm mJy$ per 24.4 $\rm kHz$ resolution. In this work, our aim is to obtain the ALFALFA 50$\%$ completeness function and compare it with \citet{Jones2015}, we thus adopt the ALFALFA 40$\%$ value \citep{Haynes2011}.} and redshift of 0 into Equation \ref{eq:S_intlim}, an SNR of 5.75 will give a close approximation to the $50\%$ completeness function of the ALFALFA survey, which can be used to derive the HIMF of the survey samples  (e.g. \citealt{Martin2010,Jones2018}). Thus, we adopt the $\rm SNR_{lim}$ of 5.75 and other technical parameters mentioned before to estimate the observed velocity integrated flux limit of CRAFTS through Equation \ref{eq:S_intlim}, from which the HI mass detection limit in terms of observed frame velocity integrated flux can be denoted by \citep{Roberts1975,Meyer2017}
\begin{eqnarray}
\label{eqn:M_lim}
\left(\frac{M_{\mathrm{HI,lim}}}{M_{\odot}}\right) = \frac{2.35 \times 10^{5}}{\left(1+{\mathrm {z}}\right)^{2}}  \left[ \frac{D_{\mathrm {L}}({\mathrm {z}})}{\mathrm {Mpc}} \right]^{2} \left(\frac{ S^{\mathrm{obs}}_{\mathrm{int,lim}}}{\mathrm{Jy\,km\,s^{-1}} } \right),
\end{eqnarray}
where $D_{\mathrm {L}}({\mathrm {z}})$ is the luminosity distance at a redshift of z.

The channel width of CRAFTS is approximately 7.6 $\mathrm{kHz}$ and the value of $S_{\mathrm{rms}}$ varies with the ZA and redshift of the source (for detail discussion, see \citealt{Zhang2019}). Figure~\ref{fig:M_lim1} shows the HI mass detection limit in the redshift and intrinsic velocity width plane at a ZA of $0\degree$ for two-pass CRAFTS. The deviations of HI mass limit at $\log{W_{\rm 50}/\mathrm{km\,s^{-1}}}=2.5$ is caused by the "standing wave", which makes it harder to detect HI galaxies with wider velocity widths.

In this paper, we use two kinds of mock catalogues to predict for CRAFTS. One is the Monte-Carlo simulation based on the ALFALFA catalogue, where no large-scale structure (LSS) of the universe is included. The low-mass end in Monte-Carlo simulation extends to $10^{6.4}\,M_{\odot}$, which can be used to test the measurements of low mass end of the HIMF and give a rather complete prediction on galaxy detection. Another is the semi-analytical simulation based on the $\mathrm{\Lambda CDM}$ model, which considers the environmental information of HI galaxies, by which we can estimate the influence of source confusion on the survey. However, the resolution of the simulation limits the low-mass end to $10^9\,M_{\odot}$.

\section{THE MONTE-CARLO SIMULATION}
\label{sec:sec3}

\subsection{The ALFALFA mass-width function }

The ALFALFA survey is the largest completed blind HI survey so far, which covers a sky region of nearly 7,000 $\mathrm{deg^2}$ to a redshift of 0.06. Here we use full ALFALFA or the $\alpha.100$ catalogue \citep{Haynes2018,Jones2018} to calculate the mass-width function (MWF; \citealt{Papastergis2011}) of the HI galaxies, which gives the distribution of the number density of HI galaxies as a function of HI mass and velocity width. The MWF of HI galaxies is an intermediate product of the calculation of the HIMF and the velocity width function (WF), which depicts the number density distribution of HI galaxies as a function of HI mass and velocity width. The HIMF and WF can be obtained by summing up the MWF along the velocity width dimension and HI mass dimension, respectively.

As blind surveys are usually flux limited, which means the samples in the survey are selected by its flux limit (for HI surveys, there is also velocity width dependence), it is required to compensate for selective effects while estimating the volume density. \cite{Schmidt1968} developed the "$1/V_{\rm max}$" method to calculate the number density of a flux-limited survey by weighting each detected source with the reciprocal of $V_{\rm max}$, where $V_{\rm max}$ is the maximal comoving volume in which the source can be detected by the survey. However, the "$1/V_{\rm max}$" method is very sensitive to the LSS of the universe, which can be reflected by the spatial distribution of clusters and voids. For nearby HI surveys, significant bias will be caused by the overdensity of HI galaxies in the local volume \citep{Martin2010}.

\begin{figure}
	\includegraphics[scale=0.36]{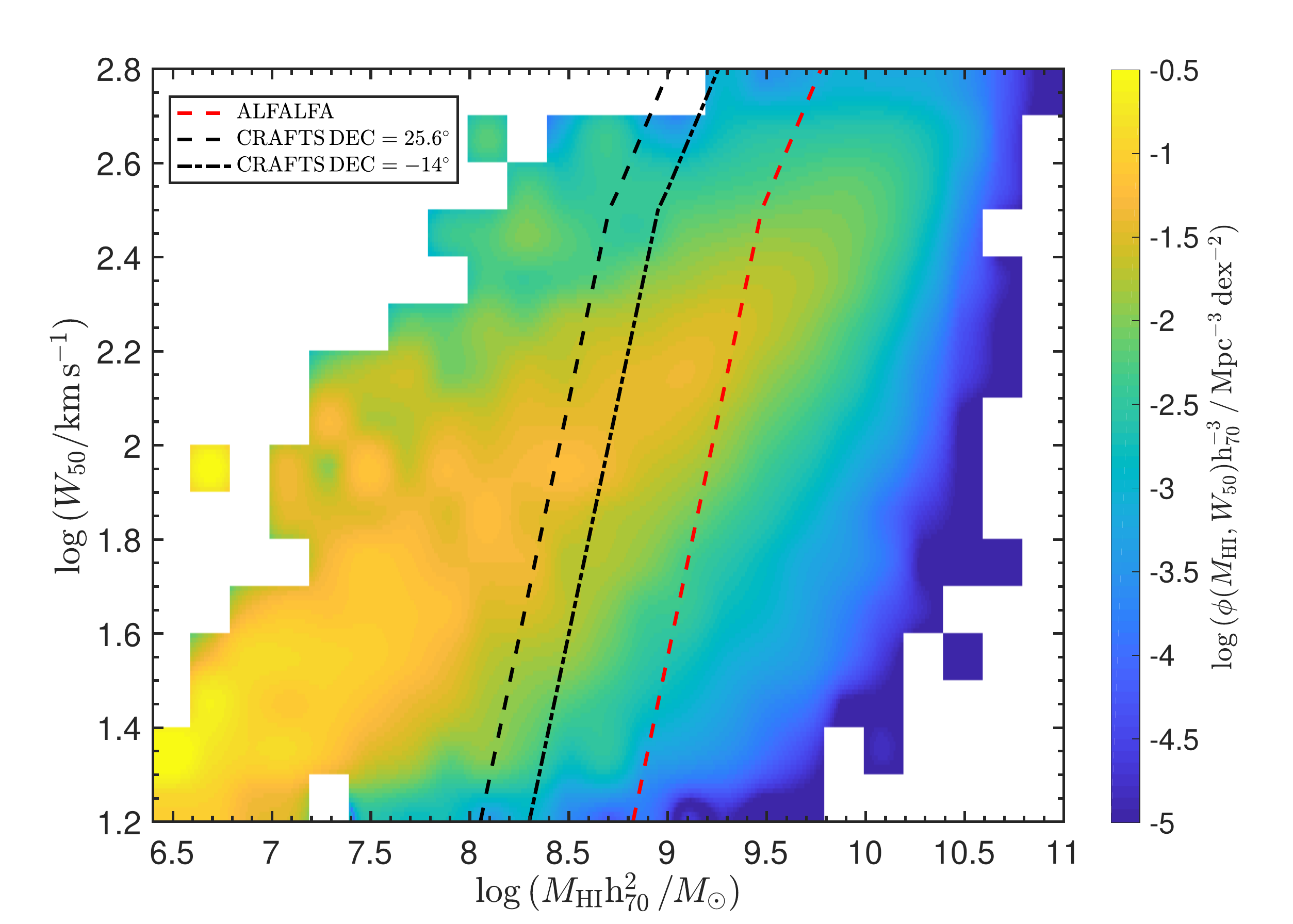}
    \caption{The mass-width function (MWF) derived from the $\mathrm{\alpha.100}$ catalogue after interpolation to a bin width of 0.01. The red dashed line represents the $50\%$ completeness HI mass limit at a redshift of 0.025 for the ALFALFA survey. Similarly the black dashed, and dash-dotted line represents the CRAFTS limit at a DEC of $25.6\degree$ and $-14.0\degree$, respectively. The galaxies located to the left of those lines will not be detected by the survey. The HI mass detection limit at a DEC of $-14.0\degree$ is higher than that at a DEC of $25.6\degree$, mainly due to the gain loss and increase of system temperature at high Zenith Angles. }
    \label{fig:MWF}
\end{figure}

The step-wise maximum likelihood (SWML) method \citep{Efstathiou1988} is developed to reduce the effect of LSS, which assumes the number density in each separated bin is constant and maximizes the joint likelihood of detecting all samples in the survey. The detection of HI galaxies will also depend on the velocity width of the galaxies. In this case, the two-dimensional SWML (2DSWML) method \citep{Loveday2000,Zwaan2003} or the "$1/V_{\mathrm{eff}}$" \citep{Zwaan2005} method is developed to calculate the MWF by splitting the MWF or $\phi$ into bins of $m=\log{(M_{\mathrm{HI}}/M_{\odot})}$ and $w=\log{(W_{\mathrm{50}}/\mathrm{km\,s^{-1}})}$. $\phi(m,w)$ can be obtained by summing up the $1/V_{\rm eff}$ of each HI galaxy in individual bin, where $V_{\rm eff}$ is the effective volume of the galaxy. $V_{\rm eff}$ can be obtained by maximizing the joint likelihood of detecting all galaxies in the survey and is equivalent to $V_{\rm max}$ in the "$1/V_{\rm max}$" method \citep{Zwaan2005}. We utilize the "$1/V_{\mathrm{eff}}$" method to calculate the MWF of the $\alpha.100$ catalogue. The detail calculation procedure of the MWF can be found in \cite{Zwaan2005,Martin2010,Papastergis2011}, and references therein.

The final samples to calculate the MWF are from the catalogue to calculate the $100\%$ ALFALFA HIMF \citep[provided by M. G. Jones via private communication]{Jones2018}. The catalogue we use here contains 22,798 HI galaxies in total with following cuts: $z \leq 0.05$, $6.4 \leq  m \leq 11$, $1.2 \leq w \leq 3.0$, and $50\%$ completeness limit for high SNR sources. The $m$ and $w$ range is chosen to make the derived number density closed to the fitted HIMF. The redshift cut is to avoid the strong contamination of the RFI from the aviation radar, the redshift gap caused by which would otherwise influence the performance of the 2DSWML method in calculating the HIMF \citep{Haynes2011}.

The MWF is first calculated in a bin width of 0.2 in $m$ and $w$ and then approximated through interpolation\footnotemark into a bin width of 0.01. The interpolated MWF derived from the $\alpha .100$ catalogue is shown in Figure~\ref{fig:MWF}. \footnotetext{The interpolation is performed through the function of "interp2" in the software of \textsc{matlab} MATLAB, the method used here is "makima", which stands for the Modified Akima cubic Hermite interpolation.}The finer distribution of $\phi(m,w)$ is used to produce the random number of $m$ and $w$ in Monte-Carlo simulation and we keep the total number density of HI galaxies a constant before and after interpolation.

%================
\begin{figure}
	\includegraphics[scale=0.38]{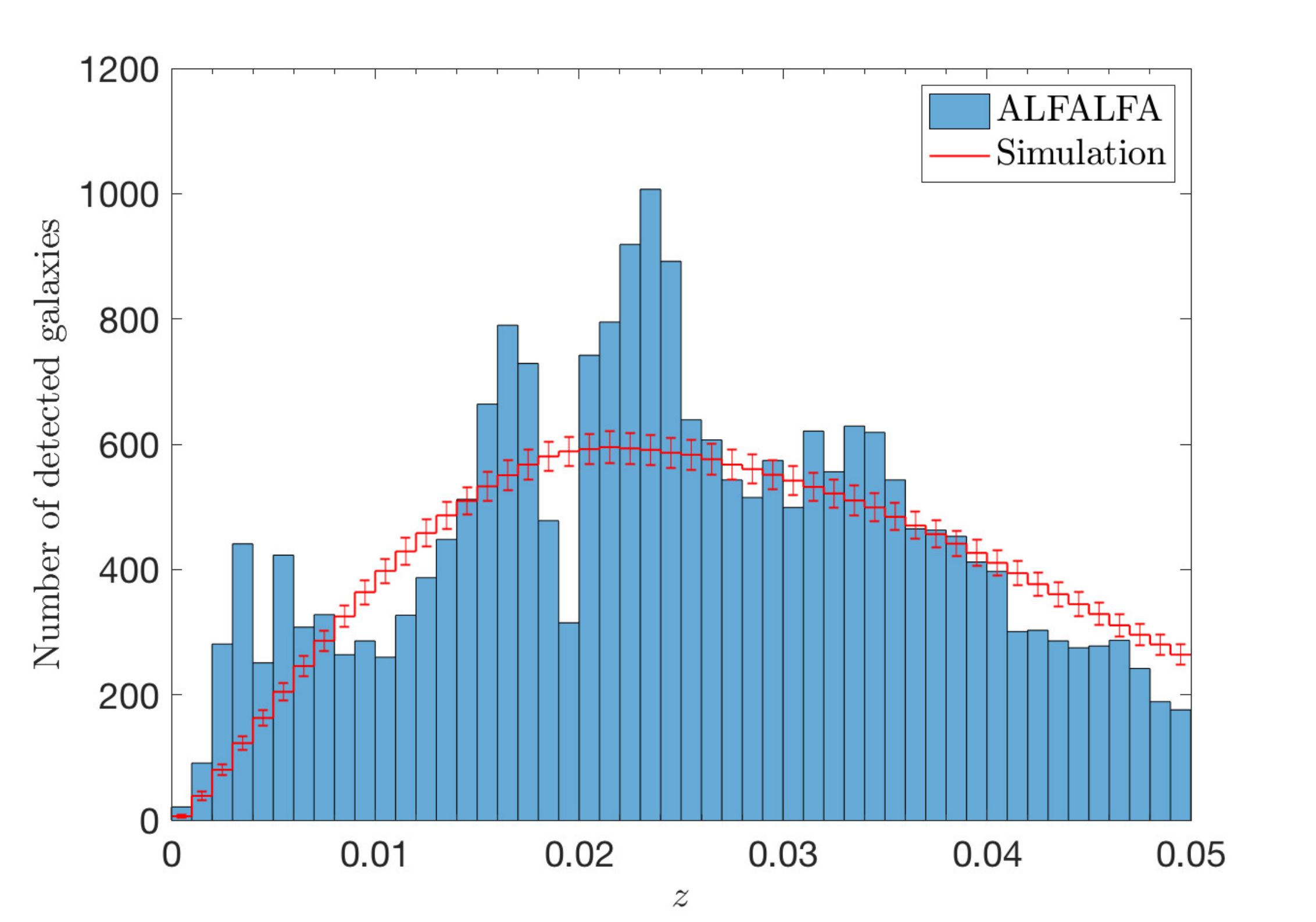}
    \caption{The redshift distribution of the $\alpha.100$ catalogue and average result from 1000 simulations, which is shown as blue histogram and red solid line, respectively. The uncertainty comes from the maximal fluctuation of 1000 simulations.}
    \label{fig:ALFALFA_num}
\end{figure}
%================

\subsection{Simulation and results}

To estimate the number of galaxies that would be detected by CRAFTS, we generate a mock catalogue through Monte-Carlo simulation, similar to the approach in \cite{Jones2015}, based on the ALFALFA MWF. We assume that HI galaxies are uniformly distributed in the universe, so that the right ascension (RA), the sine of declination (DEC) and the comoving volume ($\mathrm{V_{C}}$) of HI galaxies can be generated through uniformly distributed random numbers in CRAFTS sky. For the HI mass and velocity width, we utilize the normalized MWF mentioned before as the probability density function (PDF) of $m$ and $w$. We assume the number volume density of HI galaxies does not evolve with redshift and the value is approximately $\mathrm{ 0.1161 \,h_{70}^{-3}/Mpc^{-3} }$, which is calculated by integrating the $\alpha.100$ HIMF \citep{Jones2018} from $m$ of 6.4 to 11. The total number of HI galaxies can be obtained by multiplying the total number density of HI galaxies with the comoving volume of the survey. The redshift of a galaxy at a given comoving volume in the survey region can be approximated by its comoving distance. We assume the simulated HI galaxy can be detected when its HI mass is above the HI mass limit estimated by Equation \ref{eqn:M_lim}.

%================
 \begin{figure}
	\includegraphics[scale=0.38]{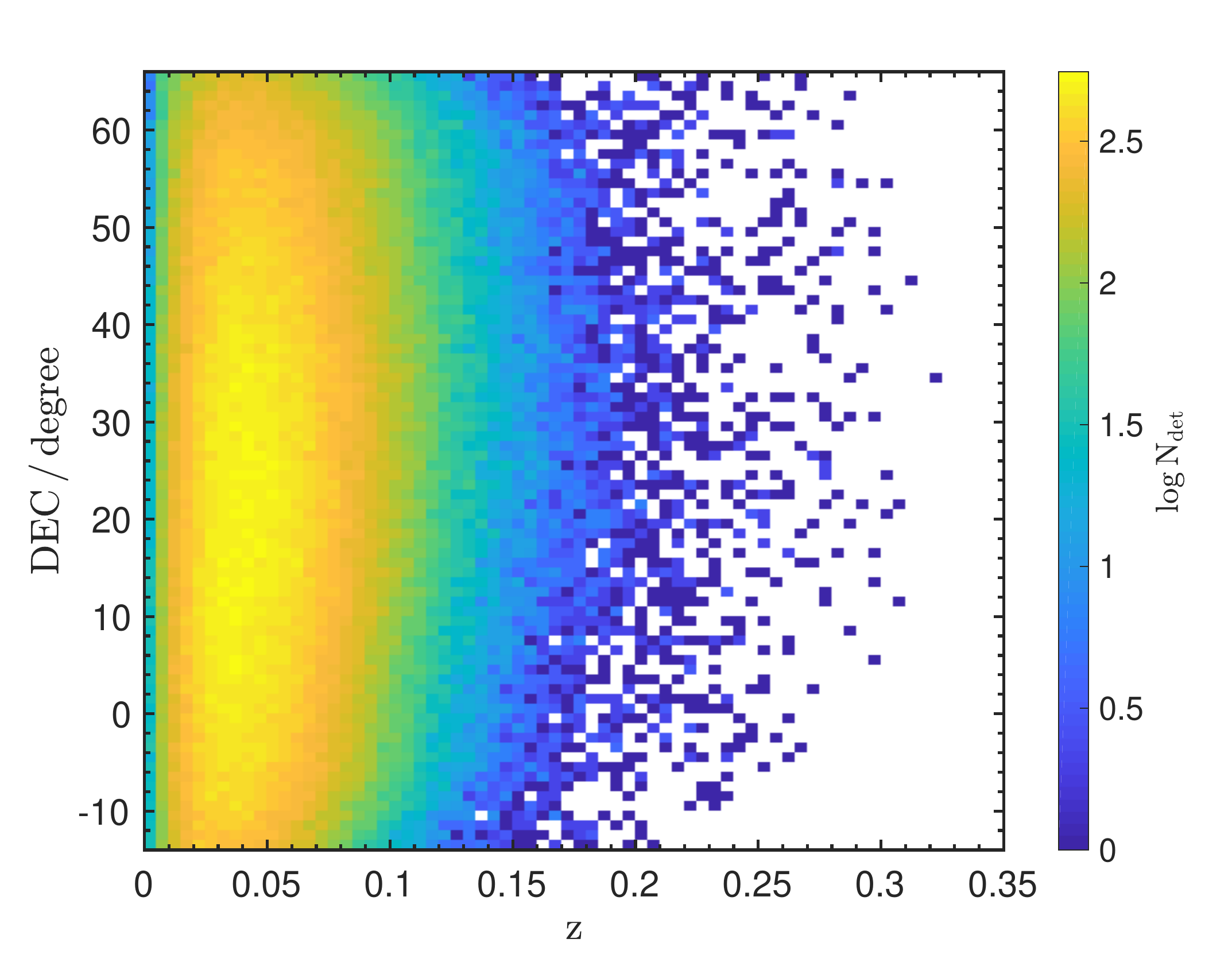}
    \caption{The number of detected HI galaxies of CRAFTS in the z-DEC plane estimated from Monte-Carlo simulation. The pixel size is 0.01 in redshift and $1\degree$ in declination. The colourbar represents the number of detected galaxies in the volume bin of a 24-h drift scan within the boundary of each pixel size. We assume the HI galaxies in the universe are uniformly distributed. The variation at different declinations implies the impact of telescope parameters on galaxy detection. }
    \label{fig:Ndet_zdec}
\end{figure}
%================

%================
 \begin{figure*}
	\includegraphics[scale=0.65]{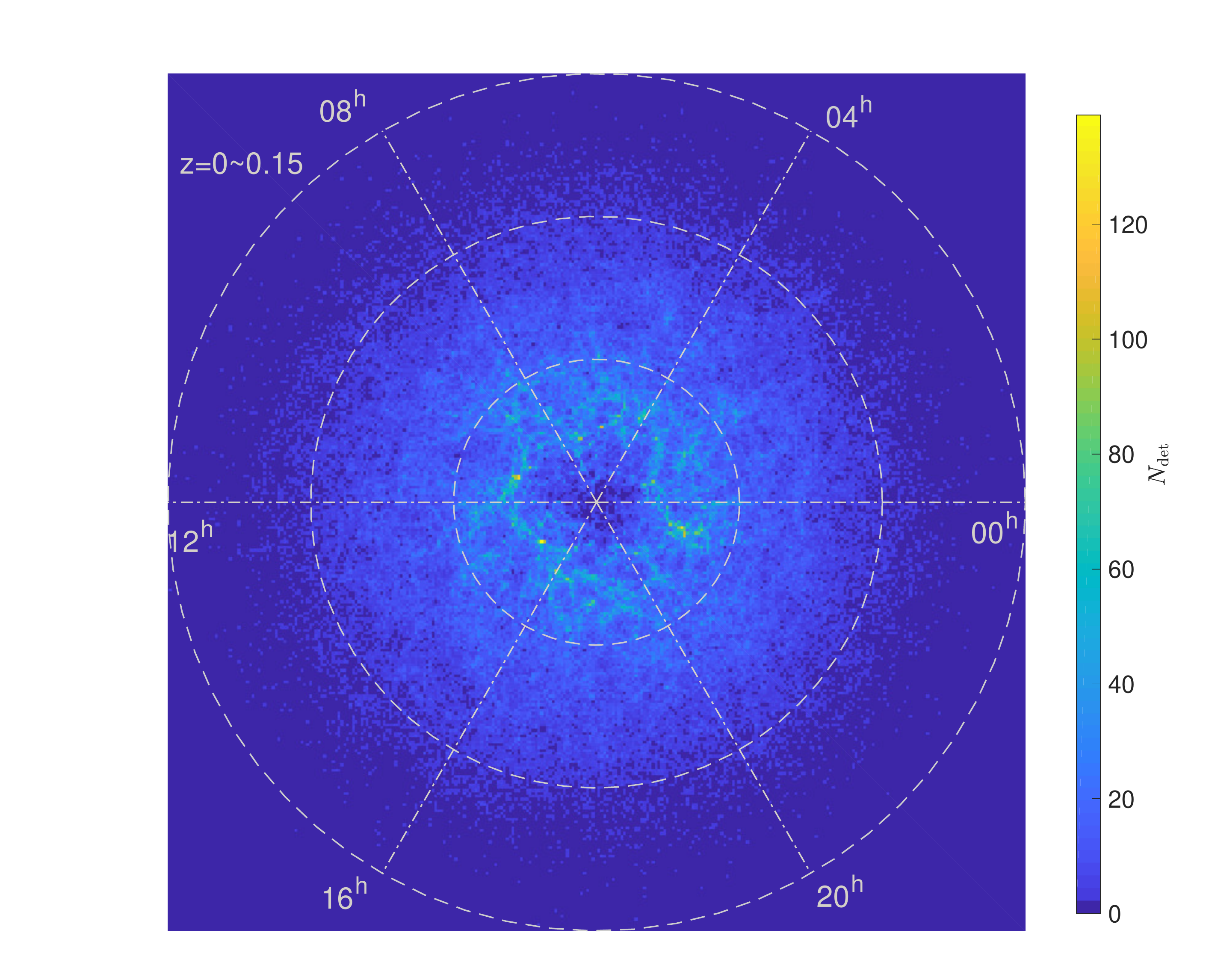}
    \caption{The number of detected galaxies projected into the z-RA plane derived from semi-analytical simulation for two-pass CRAFTS. The spacing between two adjacent gray dashed circles is 0.05 in redshift. The colourbar represents the number of galaxies detected in the area of each rectangle pixel, with a size of 0.001 in the plane. }
    \label{fig:N_zra}
\end{figure*}
%================

 %================
\begin{table*}%[t]
\begin{center}
\begin{tabular}{ c  c c c c}
\hline
Mock catalogue & \multicolumn{2}{c}{Monte-Carlo simulation} & \multicolumn{2}{c}{Semi-analytical simulation}\\ \hline
Survey strategy & {One-pass} &  {Two-pass}&  {One-pass} & {Two-pass} \\ %\hline

Number of detections & 290,000 (260,000) & 480,000 (430,000) & 250,000 &420,000 \\

Mean redshift &  0.047 (0.051) & 0.055 (0.060) & 0.052 & 0.061   \\

\hline
\end{tabular}
\caption{The predicted results of CRAFTS from Monte-Carlo and semi-analytical simulations. The Monte-Carlo simulation is based on the mass-velocity width function (MWF) of the ALFALFA survey, within an HI mass range between $\mathrm{10^{6.4}}$ and  $\mathrm{10^{11}\,M_{\odot}}$. The semi-analytic simulation is constructed in a frame of the N-body $\Lambda$CDM models with an HI mass resolution of $\mathrm{10^9\,M_{\odot}}$ \citep{Fu2013,Henriques2015,Luo2016}. We corrected the velocity widths by the ALFALFA mass-conditional velocity width function (MCWF; \citealt{Martin2010,Jones2015}) considering the discrepancy in the derived velocity width function (WF) between observations and simulations. The value in the parentheses represents the number of detected galaxies with HI masses ranging from $\mathrm{10^9}$ to $\mathrm{10^{11}\,M_{\odot}}$ in Monte-Carlo simulation.}
\label{tab:Ndet}
\end{center}
\end{table*}
%================

To test our model, we adopt the ALFALFA $50 \%$ completeness function \citep{Haynes2011} as the selection criterion to see whether we can recover the number of detected galaxies of the ALFALFA survey from Monte-Carlo simulation. Given the ALFALFA survey mainly focuses on the local volume ($z \leq 0.05$), we ignore the cosmological expansion effects. Thus, the comoving distance and luminosity distance are both approximated by $cz/H_{\rm 0}$.

The average number of detections in our 1000 simulations for ALFALFA survey is 20,984, with an average fluctuation of approximately $4.7\%$. This is approximately $8\%$ less than the observational samples, which lies above the $50\%$ completeness limit of ALFALFA and contains 22,798 (Code 1) detections in total. The comparison between simulation and observation is shown in Figure~\ref{fig:ALFALFA_num}. The neglection of the LSS in the universe might be the main reason of the deviation in redshift distribution of detected galaxies.

The FAST 19-beam receiver can reach 1.05 GHz (corresponding to $z=0.35$).  We thus consider the influence of cosmic expansion and overlook the peculiar motions of galaxies when calculating the cosmological distances. The distribution of the number of detections for CRAFTS in the z-DEC plane is shown in Figure~\ref{fig:Ndet_zdec}. CRAFTS will not be able to detect HI galaxies efficiently at redshifts above 0.1 due to the limited integration time.

\section{THE SEMI-ANALYTICAL SIMULATION}
\label{sec:sec4}

The mock catalogue we use is based on the semi-analytical models of galaxy formation from \cite{Fu2013} and \cite{Luo2016}, which is a branch of L-Galaxies models (e.g. \citealt{Kauffmann1999,Springel2001,Croton2006,DLB2007,Fu2010,Guo2011,Guo2013}). This branch includes the prescriptions of atomic gas (HI) and molecular gas ($\mathrm{H_2}$) in interstellar medium (ISM; e.g. cold gas cooling, star formation law, ram pressure stripping). The mock catalogue is based on two $\Lambda$CDM N-body simulations: the Millennium Simulation \citep{Springel2005} and the Millennium-II Simulation \citep{Boylan-Kolchin2009}, and we adopt the cosmology parameters from WMAP7 ($\mathrm{\Omega_{\Lambda}= 0.728}$, $\mathrm{\Omega_m= 0.272}$, $\mathrm{\Omega_{baryon}=0.0454}$, $\mathrm{\sigma_8= 0.807}$ and h = 0.704; \citealt{Komatsu2011}). The resolution of the simulation limits the minimum HI mass of galaxies in the mock catalogue to $\mathrm{10^9\,M_{\odot}}$. Details of the models can be found in \cite{Fu2013,Henriques2015,Luo2016} and references therein.

The WF derived from the $\Lambda$CDM simulation shows a large discrepancy compared to the ALFALFA observation results \citep{Obreschkow2009,Papastergis2011}, especially at the low-velocity width end, which is referred as the CDM overabundance problem \citep{Papastergis2011}. As discussed in \cite{Jones2015},  the predicted number of detections estimated from the ALFALFA MWF is around 60-75 percent of the results based on the $\Lambda$CDM simulation \citep{Duffy2012b} for the future survey complemented by ASKAP (the Australian SKA Pathfinder; \citealt{Johnston2008,Deboer2009}). The mock catalogue we use also overpredicts the number of HI galaxies with low-velocity widths. To make our model match with existing observation results, we use the mass-conditional velocity width function (MCWF) derived from the ALFALFA survey \citep{Jones2015} to generate the velocity widths of HI galaxies by Monte-Carlo simulation. The MCWF gives the PDF of the velocity width of a galaxy at a given HI mass, which is fitted by the Gumbel distribution \citep{Martin2010}.

%================
 \begin{figure}
	\includegraphics[scale=0.38]{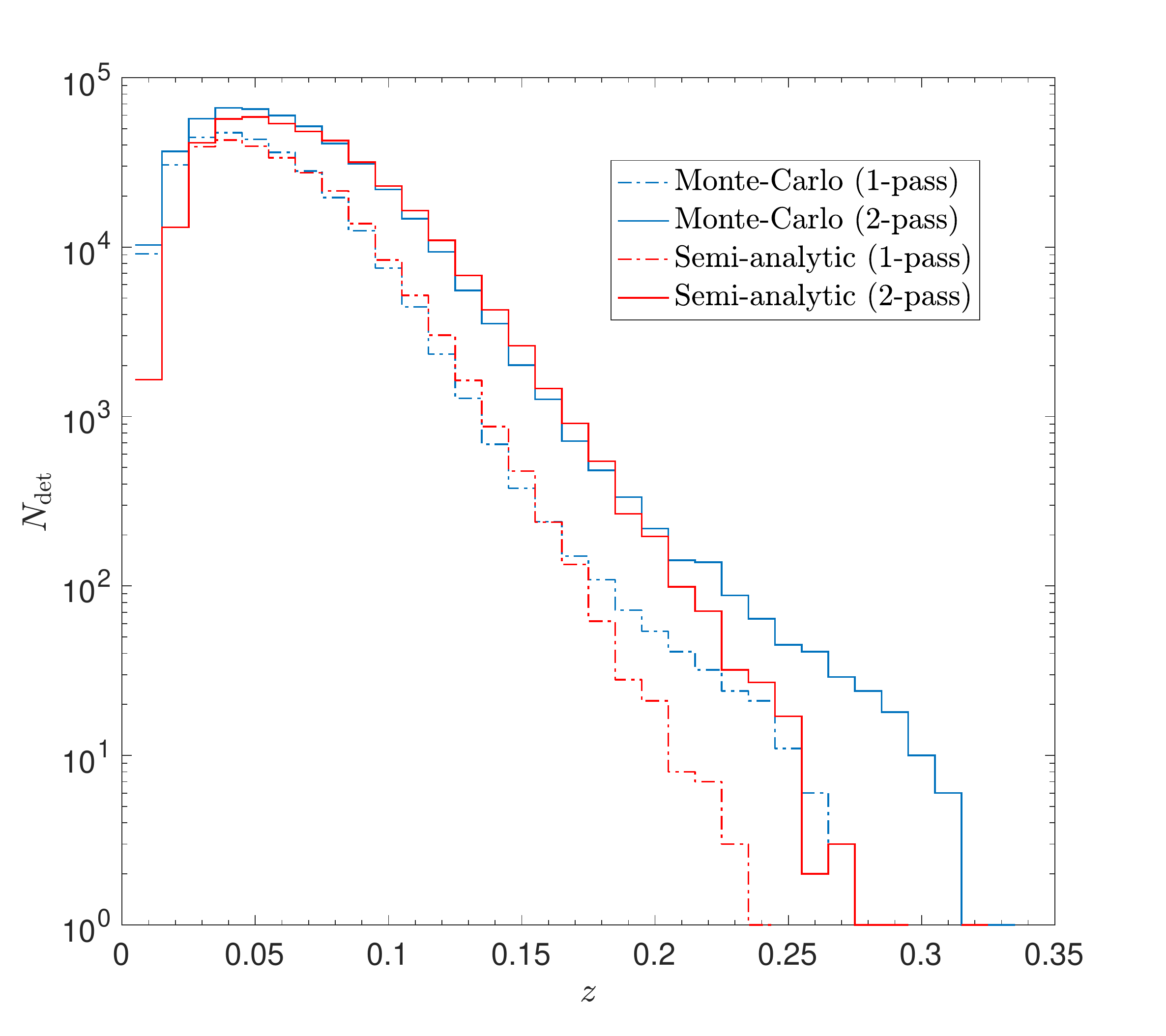}
	\includegraphics[scale=0.38]{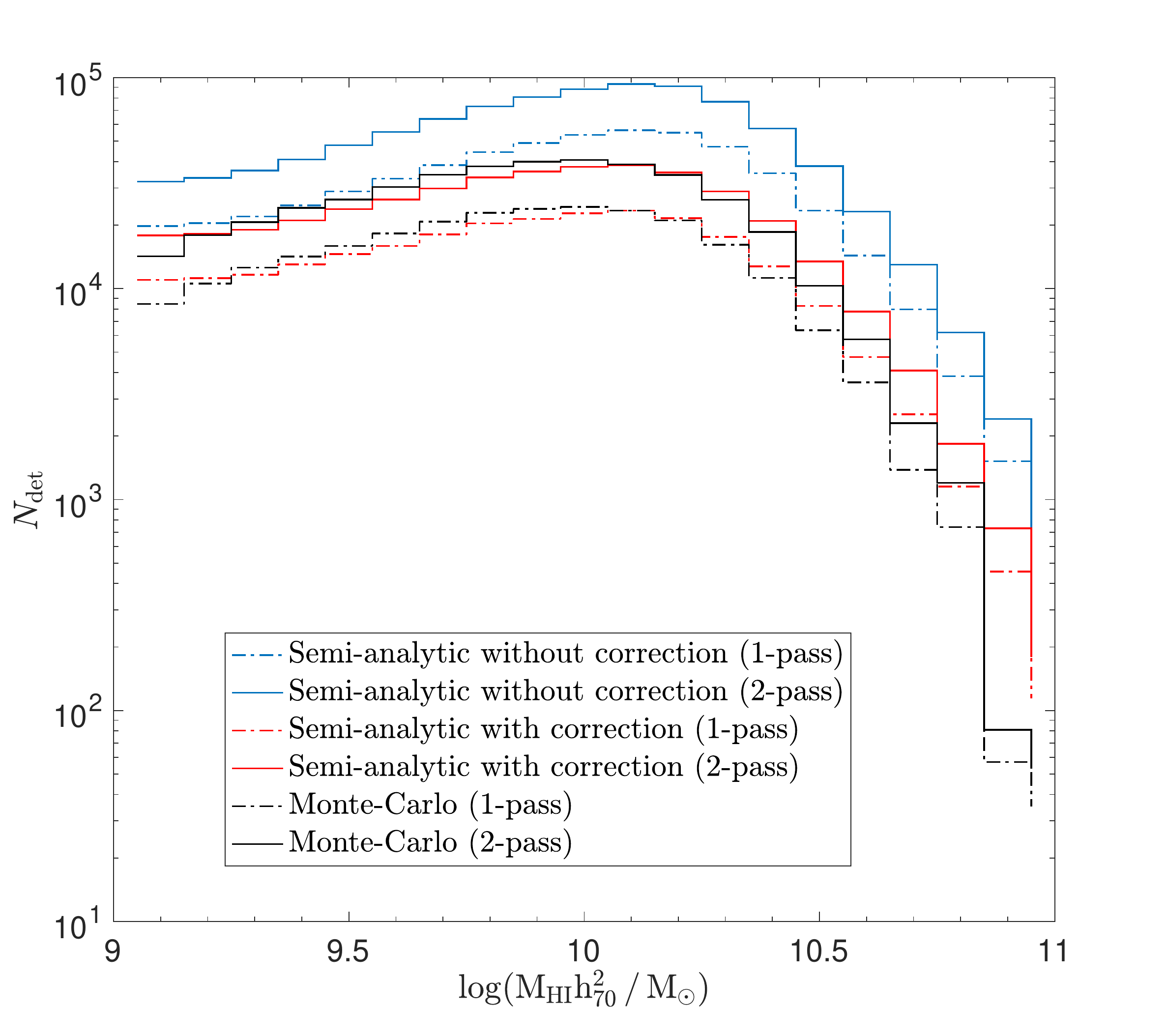}
    \caption{Top: the predicted number of galaxies detected by CRAFTS as a function of redshift. The red and blue lines represent the results from Monte-Carlo and semi-analytical simulation, respectively. Bottom: the comparison of predicted results as a function of HI mass among the semi-analytic simulation, semi-analytical simulation corrected by ALFALFA mass-conditional velocity width function (MCWF; \citealt{Jones2015}) and Monte-Carlo simulation, which are shown as blue lines, red lines and black lines, respectively. The dash-dotted lines and solid lines are from one-pass and two-pass survey, respectively.}       
    \label{fig:Ndet_z}
\end{figure}
%================

%================
  \begin{figure*}
	\includegraphics[scale=0.42]{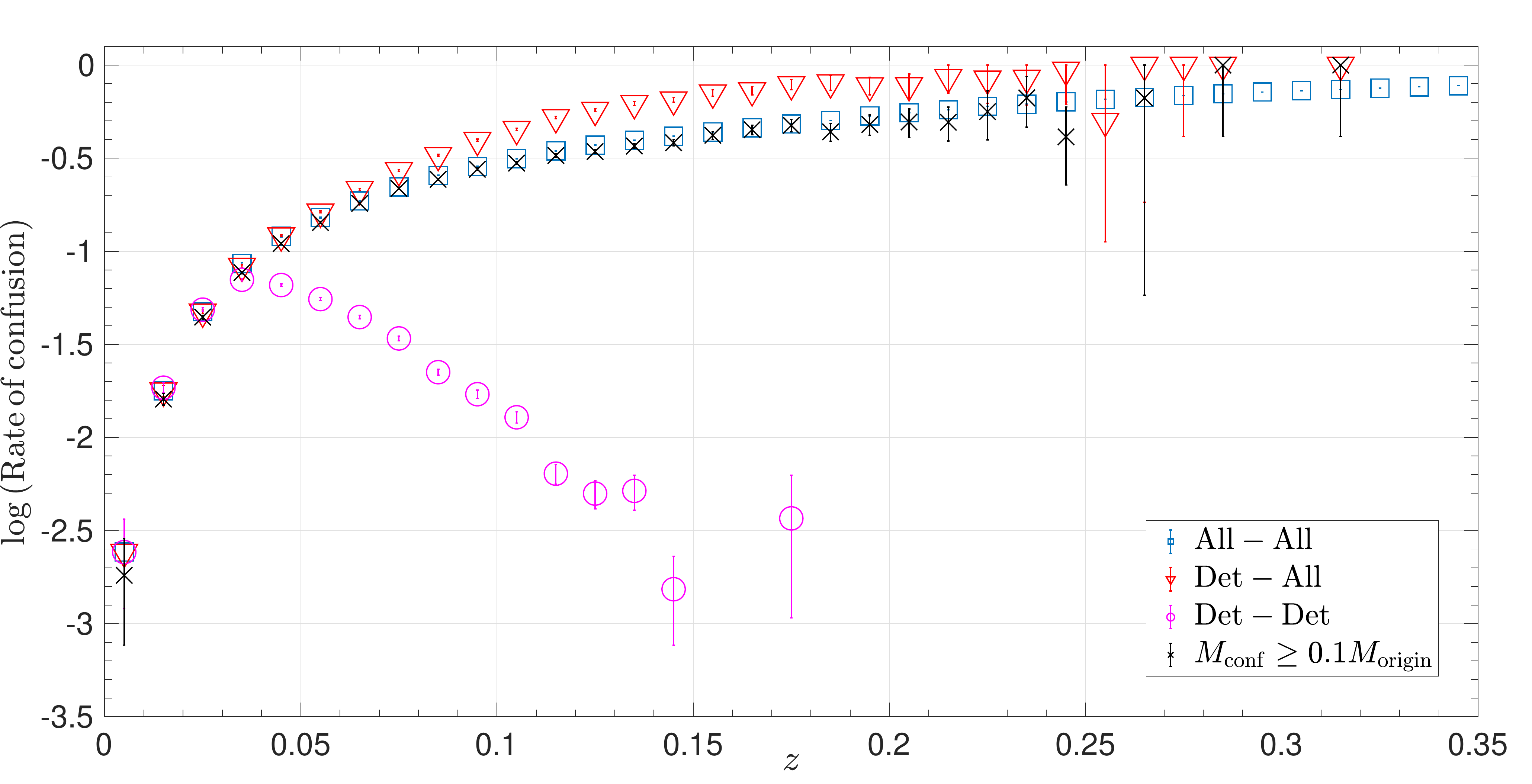}
    \caption{ The rate of confusion of CRAFTS calculated from semi-analytical simulation. The "All-All" represents the rate of confusion of simulated galaxies blended with any other simulated galaxies (blue rectangles). The "Det-All" represents the rate of confusion of detected galaxies blended with other simulated galaxies (red inverted triangles). The "Det-Det" represents the rate of confusion between detections (magenta circles). The rate of confusion when the additional confused mass $\mathrm{ M_{conf}}$ is beyond $10\%$ of its original mass $\mathrm{M_{origin}}$ is shown as black crosses. The uncertainty is from the Poisson counting error.  }
    \label{fig:cr_rand}
\end{figure*}
%================

Figure~\ref{fig:N_zra} displays the number distribution of detections projected into z-RA plane derived from semi-analytical simulation. Cosmic web features can be seen below a redshift of 0.05, which demonstrates the potential of CRAFTS in exploring the galaxy interactions in different environments and the spatial distribution of HI galaxies in the local universe. The comparison of the total number of detections and mean redshift from two simulations is presented in Table~\ref{tab:Ndet}. The predicted results from two simulations are consistent within an HI mass range between $10^{9}$ and $\mathrm{10^{11}\,M_{\odot}}$, that is because we utilize the MCWF from the ALFALFA survey to produce the velocity widths of galaxies in the semi-analytical simulation. Figure~\ref{fig:Ndet_z} illustrates the number of detections of CRAFTS as a function of redshift and HI mass in two simulations. Corrected by the ALFALFA MWF, the number of detections from semi-analytical simulation is nearly $50\%$ of the predicted results without correction. The bottom panel of Figure~\ref{fig:Ndet_z} compared the number of detections with and without correction as a function of HI mass from $10^9$ to $10^{11}\, \rm M_{\odot}$, which highlights the importance of better measurements of the WF and in matching the models to observations not just in terms of mass or luminosity function, but also the WF.

 In our predictions, one-pass CRAFTS will be able to detect nearly $2.9\times 10^5$ HI galaxies with a mean redshift of 0.047, while for two-pass survey, around $4.8\times 10^5$ galaxies will be detected with a mean redshift of 0.055. We neglect the influence of RFI on source detections. The potential impact of RFI on CRAFTS is discussed in Section~\ref{sec:sec_rfi}. The accurate number of detected galaxies from these two simulations could be different given simulations could not perfectly simulate the observed HIMF in the ALFALFA survey. However, this deviation is much smaller compared to the results without correction.

Our predicted results show that the number of HI galaxies of two-pass CRAFTS is nearly 21 times of the ALFALFA catalogue we use in this paper. That is mainly benefitted by larger sky coverage, wider bandwidth, and better sensitivity of CRAFTS. The sky coverage of CRAFTS is nearly 23,000 $\mathrm{deg^2}$, while the ALFALFA catalogue we use covers approximately 6,600 $\mathrm{deg^2}$. The ALFALFA catalogue reaches a redshift of 0.05, while CRAFTS will be able to detect galaxies efficiently to a redshift of 0.10. Our calculation indicates that two-pass CRAFTS will detect nearly 65,000 galaxies in the survey volume of ALFALFA catalogue, which is $\thicksim$3 times of the ALFALFA catalogue. The expected average sensitivity of CRAFTS is approximately 1.67 mJy per 7.6 kHz, which is $\thicksim$3.0 times of the ALFALFA survey (2.8 $\mathrm{mJy}$ per 24.4 $\mathrm{kHz}$; \citealt{Haynes2018}) after smoothing to same channel width.

\section{SOURCE CONFUSION}
 \label{sec:sec6}

Source confusion occurs when multiple sources are located within the same beam at similar velocity channels and cannot be distinguished. For HI galaxy surveys, two galaxies are confused when they are projected within one beam in the sky and overlapped in the line-of-sight direction by their velocity widths (e.g. \citealt{Duffy2012a, Duffy2012b, Jones2015}). In a shallow survey like CRAFTS, confusion effects may manifest at higher redshifts due to a large beam size at L-band and broader velocity widths of galaxies at high $z$. This may cause an overestimation of HI mass and underestimation of the number of detectable galaxies presented in a beam, which might bias the observed HIMF and WF. 

The actual source confusion relies on the distribution of HI mass, velocity width and clustering property. As the Monte-Carlo simulation does not contain any clustering property of HI galaxies, which will underestimate the rate of confusion, we only calculate the results from semi-analytical simulation. Our semi-analytical models are based on N-body cosmological simulations, which can reflect the spatial distribution of galaxies/clusters in real universe. In our previous papers of L-Galaxies SAM (e.g. \citealt{Fu2017,Henriques2020}), our results of HIMF can approximately fit the results of ALFALFA, and the models can also reproduce the 2PCF when we consider the process of stripping of the cold gas by ram pressure \citep{Luo2016}. The HI velocity width function in our mock catalogue is corrected by ALFALFA observational results. Our predictions about source confusion could act as a reference for the future survey.

%================
\begin{figure*}
	\includegraphics[scale=0.45]{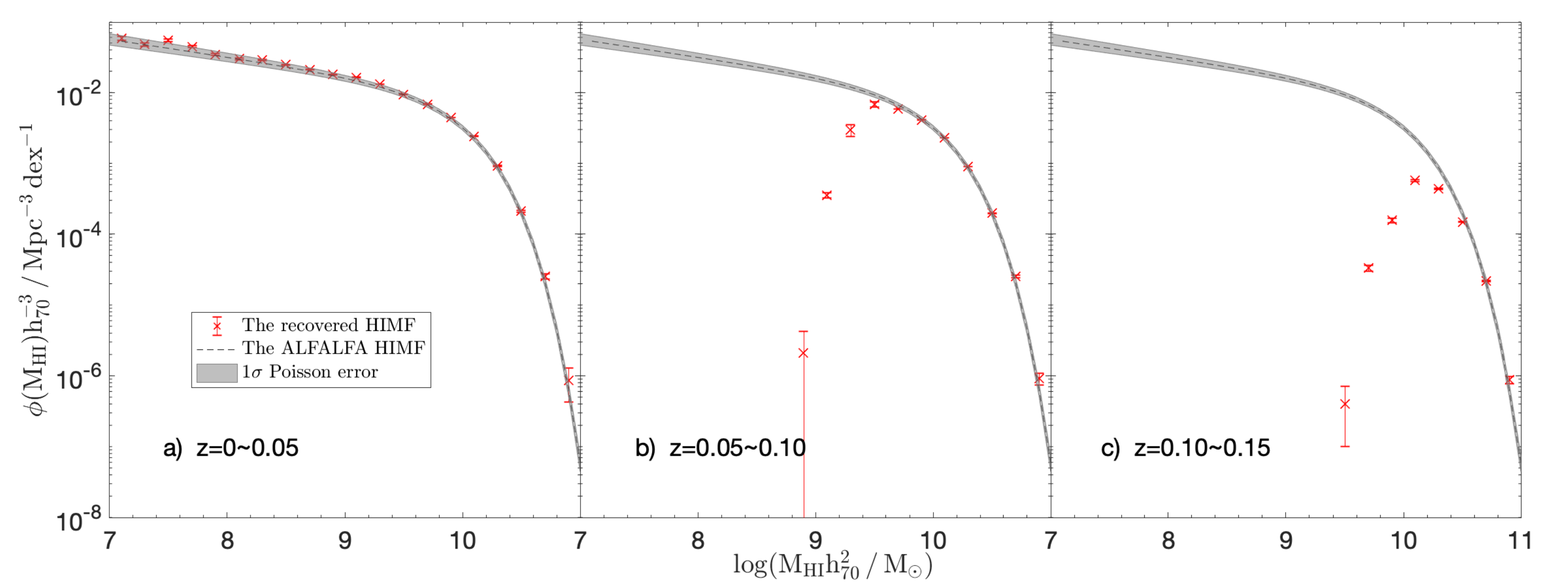}
	\includegraphics[scale=0.45]{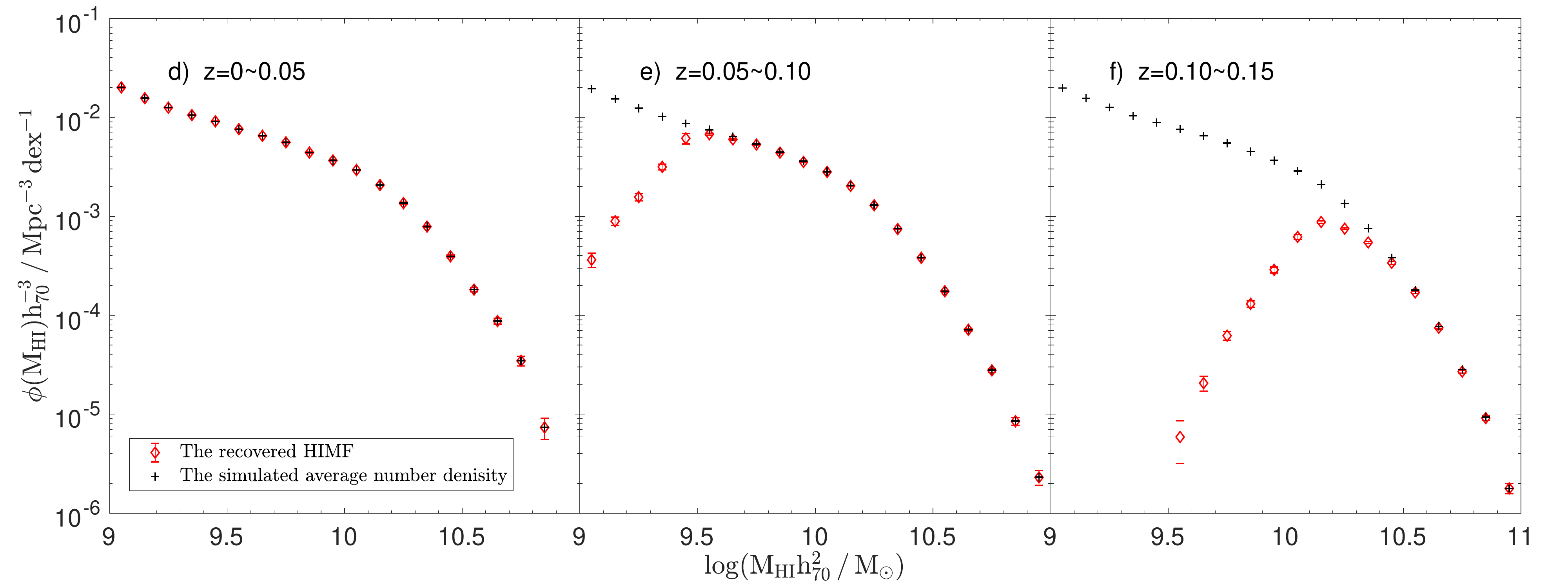}
    \caption{The recovered HIMF of CRAFTS calculated through the "$\mathrm{ 1/V_{eff}    }$" method, the uncertainty is Poissonian. Top three panels: The recovered HIMF from the Monte-Carlo simulation based on the ALFALFA MWF, the red crosses represent the observed HIMF, the ALFALFA $100\%$ HIMF \citep{Jones2018} is represented as black dashed line, with 1$\mathrm{\sigma}$ Poisson error shown as gray shaded area. Bottom three panels: The HIMF recovered from semi-analytical simulation. The red diamonds stand for the observed HIMF and the black pluses show the average number density of galaxies in the mock catalogue.    }
    \label{fig:HIMF_zslice}
\end{figure*}
%================

We followed the method in \cite{Jones2015} to define whether two galaxies are confused. To judge whether two galaxies are overlapped in the line-of-sight direction, we use the simulated channel number of the observed galaxies in the backend. We convert the redshift and velocity width of a galaxy to the channel number that the galaxy occupies in the backend of FAST, which is obtained by dividing the observed velocity width by the channel velocity width. The channel velocity width can be approximated through Equation~\ref{eq:V_ch}, we assume the channel velocity width is constant for each galaxy. Two galaxies are considered to be confused if the occupied channels have any overlaps (thus overlap in velocity space even just by the edges of flat-top profiles) and their separation in the sky is less than the beam size. The actual source confusion rate might be improved as CRAFTS will do a super-Nyquist sampling so that some confused galaxies can be actually recognized after post-survey reductions.

The rate of confusion is obtained by dividing the number of blended sources by the total number of sources in different redshift ranges with an interval of 0.01. Following \cite{Duffy2012b}, we calculate the rate of confusion in three different cases. The "All-All" case illustrates the confusion between any  simulated galaxies, no matter they are detected or not, which is shown as blue rectangles in Figure~\ref{fig:cr_rand}. The "Det-All" case represents the detected galaxies blended with any other simulated galaxies, normalized by the total number of detected galaxies, which is shown as red inverted triangles in Figure~\ref{fig:cr_rand}. The "Det-Det" case stands for the detected galaxies blended with other detections, which is shown as magenta circles in Figure~\ref{fig:cr_rand}. We also calculate the rate of confusion when the additional HI mass caused by confusion is beyond $10\%$ of its original mass, which is shown as black crosses in Figure~\ref{fig:cr_rand}. The uncertainty in Figure~\ref{fig:cr_rand} is Poissonian.

 %================
  \begin{figure*}
	\includegraphics[scale=0.265]{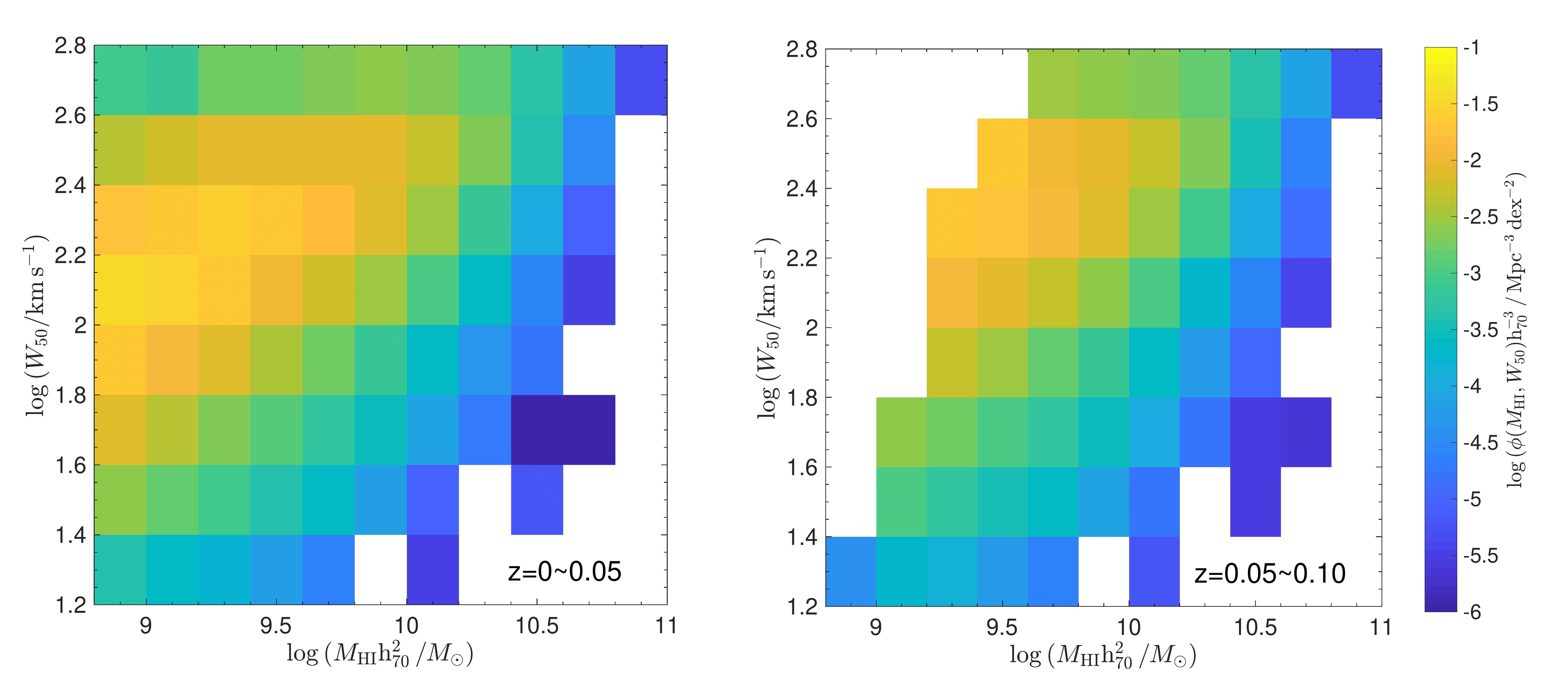}
    \caption{ The MWF revealed by CRAFTS from Monte-Carlo simulation at two redshift slices. Limited by its sensitivity, low-mass galaxies with wide velocity widths cannot be detected by CRAFTS at higher redshifts and will not be included in MWF calculation, which can be reflected by the blank area in the right-hand panel. This will cause the depression of HIMF measurements at low masses in Figure~\ref{fig:cr_rand}. } 
    \label{fig:MWF_zslice}
\end{figure*}
%================

A noticeable feature in Figure~\ref{fig:cr_rand} is that "Det-All" is higher than "All-All" at same redshifts, which indicates that the detected galaxies are more likely to be blended with other galaxies. This result is consistent with \cite{Duffy2012b}, who interpreted the effect mainly as the result of that high-mass galaxies tend to be located in denser environments and the clustering property is overpredicted by the simulation. However, we have found a similar blending feature in Monte-Carlo simulation, where no clustering property is included. Our mock catalogue has been corrected by the ALFALFA MCWF, as shown in Figure~\ref{fig:MWF}, most of low-velocity width galaxies are located in the "faint end" of the HIMF, which are hard to detect and occupy a large fraction of number densities. So the velocity widths of detected galaxies are wider than most of galaxies in the mock catalogue, which makes the detected galaxies more likely to be blended with other galaxies in the line-of-sight direction.

In \cite{Jones2016}, they estimated the expected average confused mass for FAST would be $\thicksim$$10^9\, \rm M_{\odot}$ by a redshift of $\thicksim$0.1 by assuming a cylinder volume with a velocity width of 600 $\mathrm{km \, s^{-1}}$. The average confused mass in our mock catalogue around the redshift of 0.1 is nearly $2.0\times10^9 \, \rm M_{\odot}$, which is consistent with their results. Our prediction implies that "Det-Det" will reduce nearly $2\%$ of detected galaxies in semi-analytical simulation, which suggests that source confusion will not play a significant role in the detection of HI galaxies for CRAFTS.

\section{THE RECOVERY OF THE HIMF}
\label{sec:sec5}

The HIMF depicts the number density of HI galaxies as a function of HI mass, which can be well fitted by the Schechter function
\citep{Schechter1976}. The HIMF could be expressed in the form of
\begin{eqnarray}
\label{eq:HIMF}
\phi(M_{\rm HI}) &=& \frac{dN_{\rm gal}}  { dV \, \log {M_{\rm HI} } } \,, \\
	&=& {\rm ln}(10)\, \phi_{*}\, \left(\frac{M_{\rm HI}}{M_{*}}\right)^{\alpha +1}\, \mathrm{e}^{-\left(\frac{M_{\rm HI}}{M_{*}}\right)}\, ,
\end{eqnarray}
where $dN_{\rm gal}$ is the average number of galaxies in comoving volume element $dV$, $\phi_{*}$ is the normalization constant, $M_{*}$ is the "knee mass" (we will often use $\mathrm{m_{*}}\,=\,\log{M_{*}/M_{\odot}}$), and $(\alpha+1)$ is the low-mass slope, which is usually referred as the "faint end" slope of the HIMF.

Measuring the HIMF is one of main science goals of CRAFTS. As discussed in Zh19, beam size of FAST is slightly small at low frequencies to adjust the wide bandwidth of the receiver, which will add complexity to the calculation of the simulated integration time and selection function.  For example, at ZAs below $26.4\degree$, the actual beam size of FAST at a frequency of 1050 MHz will be nearly 0.11 arcmin smaller than the theoretical beam size if we assume beam size is inversely proportional to frequency.

To simplify the calculation of the HIMF, we assume that the beam size of FAST is inversely proportional to frequency, we change the parameter value of equation (1) and (2) in Zh19 to $\mathrm{m1=3.24\, arcmin}$ when ZA is below $26.4\degree$ and $\mathrm{m2=0}$, where m1 is the beam size at a frequency of 1250 MHz and m2 is the correction parameter. This will slightly decrease the sensitivity of the survey but give a more simplified form of the simulated selection function.

We use the "$\mathrm{1/V_{eff}}$" method mentioned previously to calculate the HIMF from the selected samples of two-pass CRAFTS. The survey volume is separated into three parts from a redshift of 0 to 0.15 with an interval of 0.05. The recovered HIMF from Monte-Carlo and Semi-analytical simulation is shown in Figure~\ref{fig:HIMF_zslice}. As in Monte-Carlo simulation, CRAFTS is not sensitive enough to fully recover the number density of galaxies with $m$ below 7, we calculate the HIMF from $m$ of 7 to 11 with a bin size of 0.2, same as the bin size to calculate the ALFALFA HIMF. The result is shown in the top three panels of Figure~\ref{fig:HIMF_zslice}. The HIMF from semi-analytical simulation, limited by its mass resolution, is obtained from $m$ of 9 to 11 with an interval of 0.1. The comparison of the recovered HIMF and average number density from semi-analytical simulation is presented in the bottom three panels of Figure~\ref{fig:HIMF_zslice}. From Figure~\ref{fig:HIMF_zslice} we can see that CRAFTS will be able to recover the "faint end" of the HIMF to $ \mathrm{ 10^{7}\, M_{\odot}  }$ and measure the "knee mass" to a redshift of 0.1, which will be the first time to achieve such a deep observation of the global "knee mass". 

Figure~\ref{fig:MWF_zslice} illustrates the measured MWF by CRAFTS from Monte-Carlo simulation at $z$=0$-$0.05 and 0.05$-$0.10. At higher redshifts and low HI masses, CRAFTS is not sensitive to galaxies with wide HI profiles. These galaxies are not detected, so cannot be included in HIMF calculation, which can be reflected by the blank area at $z$=0.05$-$0.10. The low-mass end of the recovered HIMF therefore underestimates the true HIMF at $z>0.05$, which can be seen from Figure~\ref{fig:HIMF_zslice}. For a single-dish telescope like FAST, the measurement of the HIMF might be biased by source confusion, which will be discussed in Section~\ref{sec:sec7.2}.

\section{THE NORMALIZATION OF THE HIMF}

\label{sec:sec9}

In 2DSWML, the non-normalized density parameter can be obtained from the iteration of 
\begin{eqnarray}
\phi_{jk}=\frac{n_{jk}}{\sum_{i} \frac{H_{ijk}}{\sum_{m}\sum_{n}H_{imn}\phi_{mn}  }},
\label{eqn:phi_jk}
\end{eqnarray}
where $\phi_{jk}$ is the non-normalized density parameter in $m$ bin of $j$ and $w$ bin of $k$, $n_{jk}$ is the total number count of galaxies in bin $jk$ and $H_{ijk}$ is the fraction of areas of bin $jk$ in $(m,w)$ plane that lies above the selection function of galaxy $i$. The detail form of $H_{ijk}$ can be found in the appendix of \cite{Martin2010,Papastergis2013} and references therein.

The denominator of Equation~\ref{eqn:phi_jk} represents the total number of galaxies located in bin $jk$ in the survey volume, which is corrected by the selection function of each galaxy in the sample \citep{Zwaan2005,Papastergis2013}. Assuming the number density of galaxies is constant, the denominator of Equation~\ref{eqn:phi_jk} is proportional to the survey volume and can be treated as the "effective volume". Similar to the standard "$1/V_{\rm max}$" method, Equation~\ref{eqn:phi_jk} can be converted in the form of summing the reciprocal of the "effective volume" of each galaxy in bin $jk$ \citep{Zwaan2005}.

%================
\begin{figure}
	\includegraphics[scale=0.35]{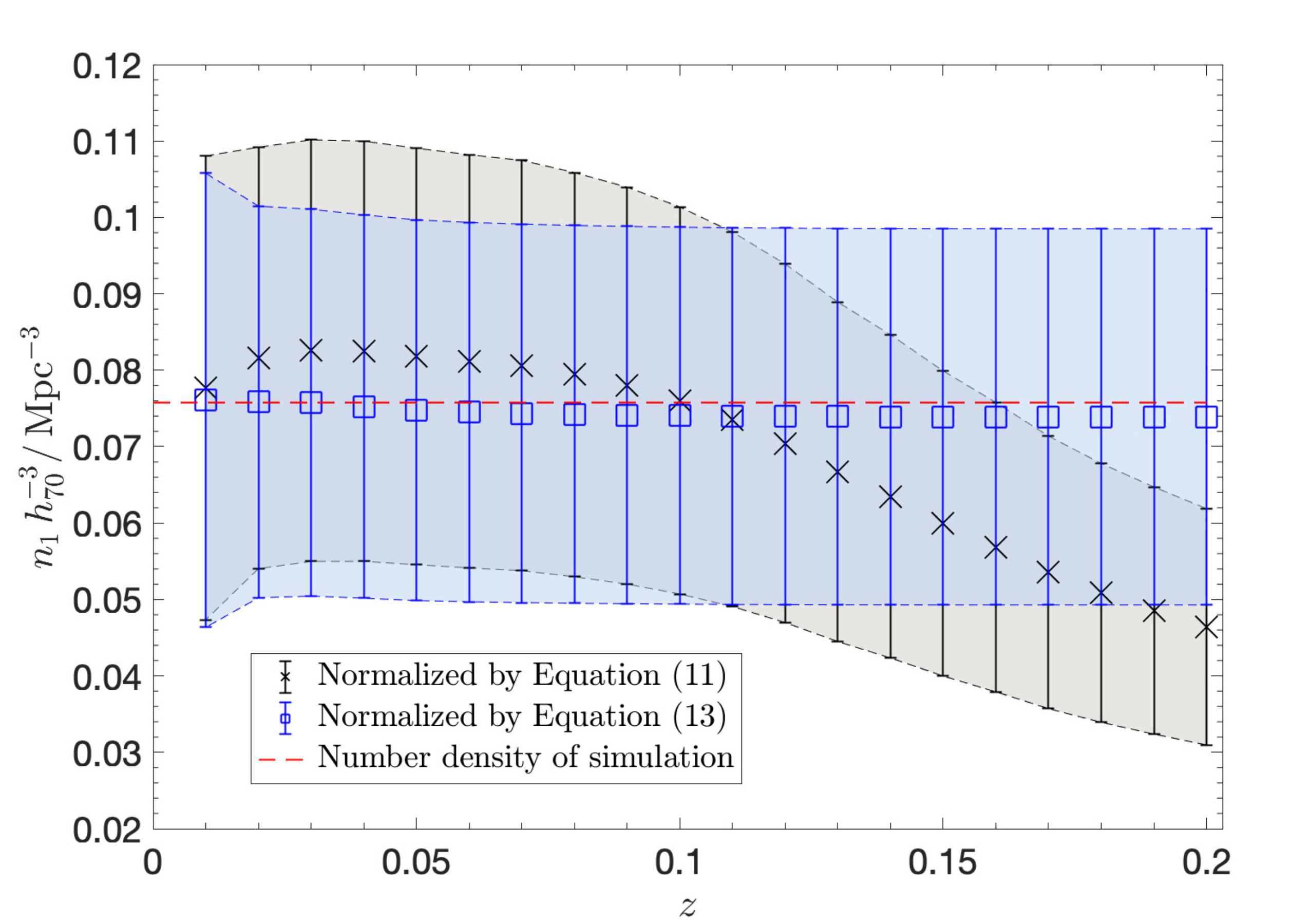}
    \caption{The average number density in different volumes derived from CRAFTS samples in Monte-Carlo simulation. The horizontal axis represents the redshift boundary of survey volume. We make total number density of galaxies constant in the simulation, which is shown as red dashed line. The black crosses are the results calculated by Equation~\ref{eqn:n_1} \citep{Martin2010}, we can see the fluctuation trend of $n_{1}$ is similar to that of the number of detected galaxies as a function of redshift, which may imply that the variation is caused by the non-uniform distance distribution of detected galaxies. The blue rectangles represent the results calculated by Equation~\ref{eqn:n_1jk}, which looks more stable. The uncertainty is Poissonian. }
    \label{fig:n_mean}
\end{figure}
%================

The normalization of the HIMF is obtained by matching the total number density to the estimator of average number density. There are three estimators described in \cite{Davis1982}, which are most suitable in different situations. As the selection function of the ALFALFA is better understood, the estimator which is less prone to bias is adopted, which can be defined as \citep{Davis1982,Martin2010}
\begin{eqnarray}
n_{1}=V_{\mathrm{survey}}^{-1} \int \frac{n(D)\,d\,D}{S(D)}=V_{\mathrm{survey}}^{-1}\sum_i \frac{1}{S(D_i)},
\label{eqn:n_1}
\end{eqnarray}
where $n(D)\,d\,D$ is the number of galaxies in a spherical shell of thickness $d\,D$ and radius $D$, $V_{\mathrm{survey}}$ is the total volume of the survey and $S(D)$ is the selection function at the distance of $D$. This method assumes galaxies are uniformly distributed and considers galaxy $i$ located at the center of the corresponding shell. Thus $1/S(D_i)$ represents the total number of galaxies in the shell centered at $D_i$, where $D_i$ is the distance of galaxy $i$.

As the survey is flux-limited, the number of detected galaxies will decrease at high redshifts, which means $\sum_i 1/S(D_i)$ due to the selective effects is not proportional to the comoving volume. The distance distribution of distant galaxies is much sparser than those nearby, which will cause a variation of $n_{1}$. We use the galaxies detected by CRAFTS in Monte-Carlo simulation to calculate $n_{1}$ in different survey volumes. The result of Equation~\ref{eqn:n_1} is represented as black crosses in Figure~\ref{fig:n_mean}, from which we can see a noticeable decline of $n_1$ as survey volume increases.

%================
\begin{figure*}
	\includegraphics[scale=0.42]{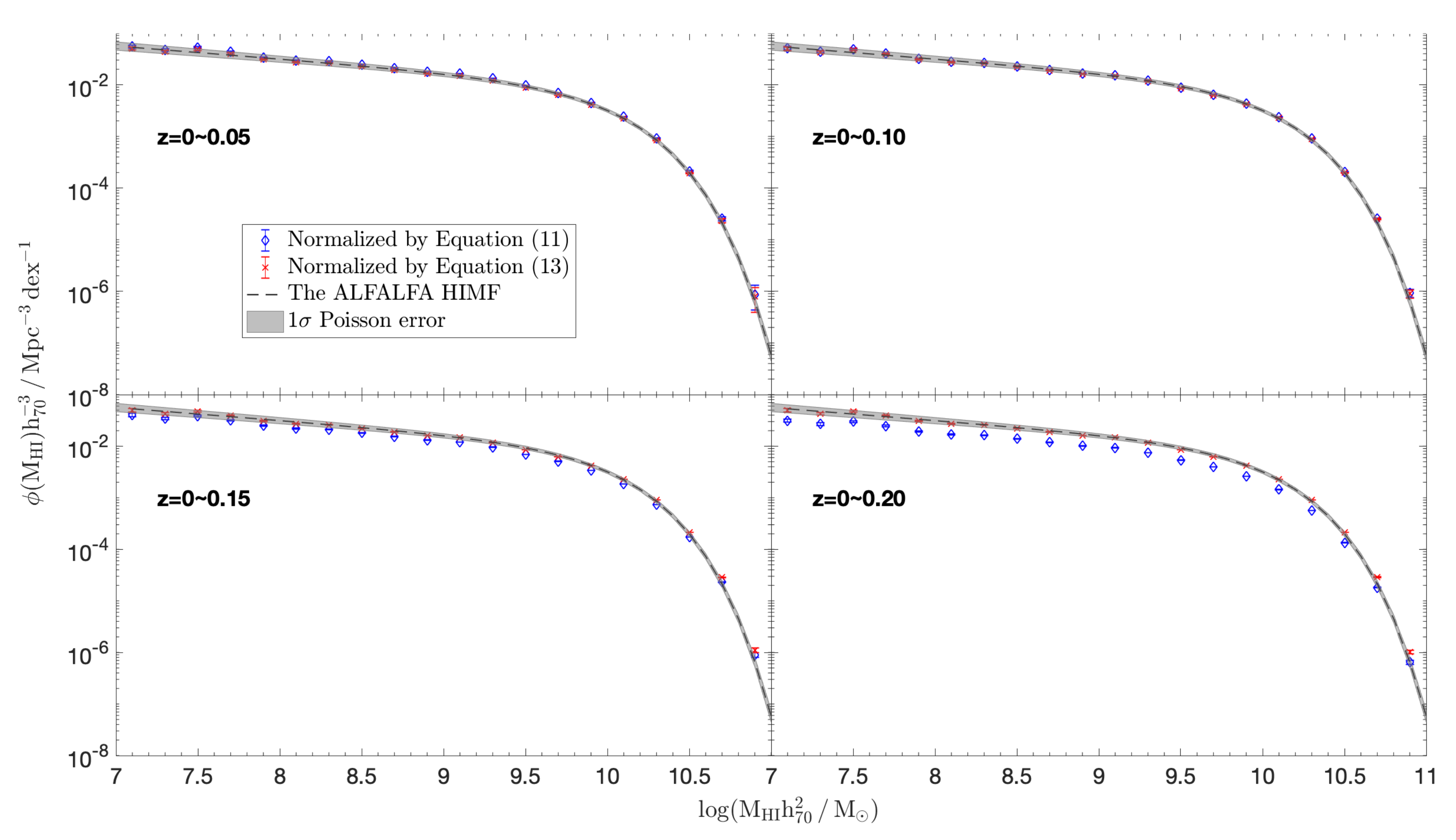}
    \caption{The comparison of HIMFs normalized by equations~\ref{eqn:n_1} \citep{Martin2010} and~\ref{eqn:n_1jk} in different redshift ranges, which is shown in blue diamonds and red crosses, respectively. The HIMFs are calculated from CRAFTS detections in Monte-Carlo simulation. The dashed line represents the ALFALFA 100$\%$ HIMF \citep{Jones2018}, with 1$\sigma$ Poisson counting error shown as gray shade area. As can be seen from the bottom two panels, when including samples from higher redshifts, the HIMF would be underestimated by Equation~\ref{eqn:n_1}. }    
    \label{fig:HIMF_zall}
\end{figure*}
%================

%================
\begin{figure*}
	\includegraphics[scale=0.38]{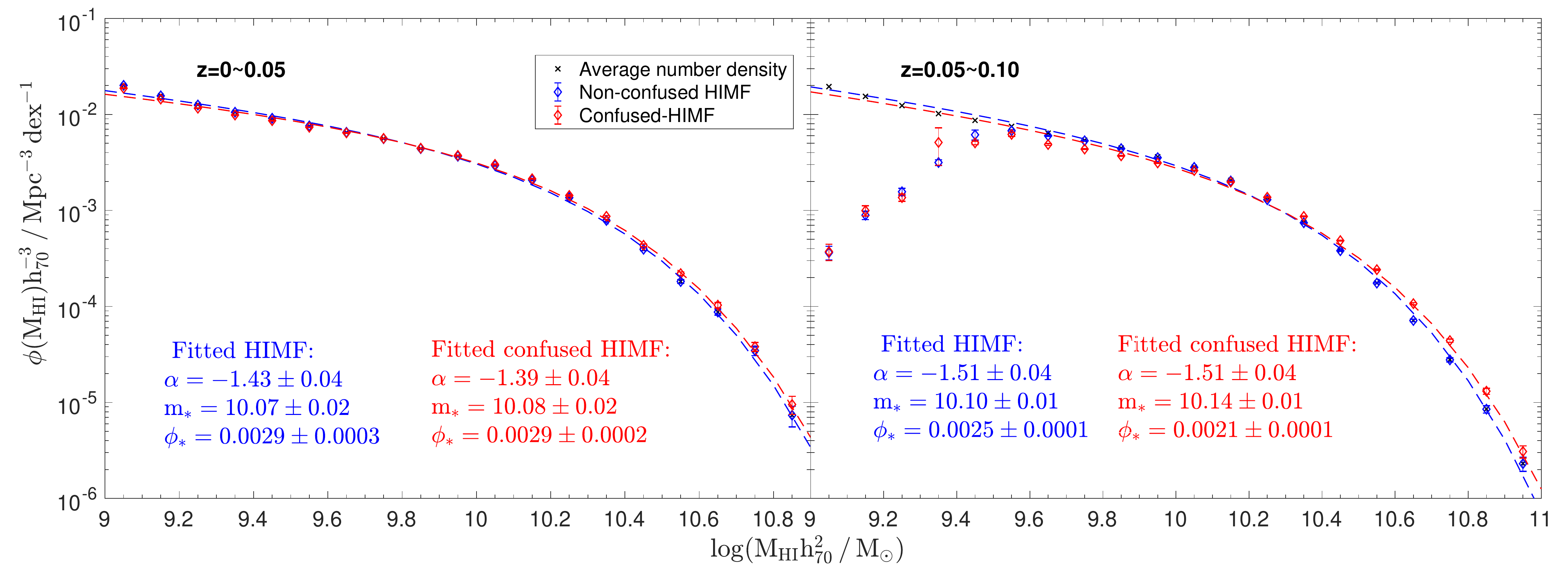}
    \caption{The comparison of confused and non-confused HIMF or the observed HIMF with and without considering source confusion effects simulated for CRAFTS. The HIMF is calculated from semi-analytical simulation by the "$\mathrm{1/V_{eff}}$" method, the uncertainty is from the Poisson counting error. The black crosses represent the average number density of the mock catalogue, the blue and red diamonds stand for the non-confused and confused HIMF, respectively. The fitted Schechter functions are denoted by dashed lines. At $z$=0.05$-$0.10, the HIMFs are fitted in an $m$ range of 9.7$-$11.0 and the "faint end" slope is adopted from the average number density of the catalogue.  }
    \label{fig:HIMF_conf}
\end{figure*}
%================

The 2DSWML method splits number density into different mass and velocity width bins. For galaxies located at same bin, 
the selective effects will not differ too much, therefore the number of detection will not be strongly biased by distance. We can also calculate the average number density in each bin and then sum up to obtain the total number density. Thus, $n_1$ can be interpreted by

\begin{eqnarray}
n_{1}=\sum_{j}\sum_{k}n_{1,jk}=\sum_{j}\sum_{k}V_{\mathrm{survey},jk}^{-1}\int \frac{n_{jk}(D)\,d\,D}{S_{jk}(D)},
\label{eqn:n_11}
\end{eqnarray}
where $n_{1,jk}$ is the average number density in bin $jk$, $V_{\mathrm{survey},jk}$ is the survey volume of bin $jk$, the boundary of which can be obtained from the most distant galaxy in bin $jk$. Similar to Equation~\ref{eqn:n_1}, $n_{jk}(D)\,d\,D$ is the number of galaxies belonging to bin $jk$ in a spherical shell of thickness $d\,D$ and radius $D$, and $S_{jk}(D)$ is the fraction of galaxies detectable in bin $jk$ at the distance of $D$. In 2DSWML, $n_{1,jk}$ can be interpreted by 

\begin{eqnarray}
n_{1,jk}=V_{\mathrm{survey},jk}^{-1} \sum_{l}\frac{1}{H_{ljk}},
\label{eqn:n_1jk}
\end{eqnarray}
for every galaxy $l$ in bin $jk$ and $H_{ljk}>0$.

The total average number density calculated from Equation~\ref{eqn:n_1jk} is shown as blue rectangles in Figure~\ref{fig:n_mean}. One can see the results are much more stable as the survey volume increases, within $5\%$ difference compared to the mean number density integrated from the input HIMF. For comparison, the maximum deviation of $n_{1}$ calculated from Equation~\ref{eqn:n_1} is nearly $10\%$ at redshifts below 0.1. Figure~\ref{fig:HIMF_zall} compares the HIMFs from Monte-Carlo simulation within four volumes normalized by equations~\ref{eqn:n_1} and~\ref{eqn:n_1jk}, from which we can see equation~\ref{eqn:n_1} would lead to an underestimation of HIMF when including sources in larger volumes.

Equation~\ref{eqn:n_1jk} normalizes the HIMF by the detections in every distinct bin, thus if there are a small number of detected galaxies in a bin, it may induce large uncertainty in the calculation of the total average number density, which will influence the accuracy of the normalization. While calculating the number density in different redshift slices, the redshifts of some galaxies are very close to the lower boundary of the redshift bin, which may underestimate the survey volume of individual bin and overestimate the total number density. Equation~\ref{eqn:n_1} normalizes the HIMF by including all of the galaxies in the survey volume and the detected galaxies in every redshift bin tend to be uniformly distributed. Thus, equation~\ref{eqn:n_1}  will give a more accurate estimation while calculating HIMFs in different redshift slices. Our results in Figure~\ref{fig:HIMF_zslice} and Figure~\ref{fig:HIMF_conf} are both calculated from equation~\ref{eqn:n_1}. These two methods could be complementary in normalizing the HIMF.

\section{DISCUSSION}
\label{sec:sec7}

\subsection{The impact of confusion effect on CRAFTS HIMF measurements }
\label{sec:sec7.2}

The measured HIMF might be biased by confusion effect from real distribution, especially at higher redshifts. The actual source confusion relies on the distribution of HI mass, velocity width, and clustering property. In this work, we use the samples from semi-analytical simulation to roughly estimate the potential influence of source confusion on the observed HIMF of CRAFTS. Following \cite{Jones2015}, we sum up the HI mass of blended sources together as the final confused mass to achieve an upper limit of the impact of confusion on HIMF measurements. For detections blended with non-detections, we assume other properties like velocity widths and positions stay constant after considering source confusion. For cases of confusion between two detections, we count blended sources as a single source, consider the boundary of blended HI profiles in the line-of-sight direction as the final confused velocity width and adopt the average value of positions. The criteria of source confusion are described in Section~\ref{sec:sec6}.

We calculate the confused HIMF to a redshift of 0.1, that is because, as illustrated in Figure~\ref{fig:HIMF_zslice}, CRAFTS will be able to measure the "knee mass" of the HIMF to that redshift. The result is shown in Figure~\ref{fig:HIMF_conf}, from which we can see larger uncertainties are produced by source confusion at higher redshift end. As mentioned previously, CRAFTS will not be able to recover the number density of low-mass galaxies at $z$=0.05$-$0.10, we thus fit the HIMF in an $m$ range from 9.7 to 11.0 by adopting the "faint end" slope of the average number density of the mock catalogue and assuming the "faint end" slope does not change for confused HIMF. The deviation of $m_{\ast}$ caused by source confusion is within 1$\sigma$ and 4$\sigma$ Poisson counting error at redshifts of $z$=0$-$0.05 and 0.05$-$0.10, respectively. As our assumptions have overestimated the confused HI mass to a large extent, the actual impact of source confusion could be significantly weaker.

\cite{Jones2015} found that confusion will steepen the "faint end" slope of the HIMF, but our fitted results from semi-analytical simulation seems to differ from their results. That is likely because the low-mass end of the HIMF we measured only reaches $\mathrm{10^9\,M_{\odot}}$, which is not the actual "faint end" of the HIMF. The "faint end" slope of the HIMF is mainly constrained by low-mass galaxies, only a few of which can be detected at very low redshifts due to the limited sensitivity of the telescope. While at low redshifts, the angular size of the beam corresponds to a smaller physical size, thus making confusion less likely. As the HI mass increases, source confusion will have a larger impact on detected galaxies. From the left-hand panel of Figure~\ref{fig:HIMF_conf}, we can see the decreased number density caused by source confusion when $m$ is below $m_{\ast}$. The decreased number density at $m<m_{\ast}$ will drag down the "faint end" slope while fitting, which is consistent with the results in \cite{Jones2015}. Considering the extremely low number density at high-mass end, a small number of added sources will lead to notable changes of observed number density, which could influence the observed "knee mass" of the HIMF.

\subsection{The RFI status at FAST site }
\label{sec:sec_rfi}

The impact of RFI on source detection can be vital on FAST, especially considering the receiver of FAST covers a wide range of frequencies. Currently, the severest contamination is from satellites and aircraft navigation beacons, which corresponds to redshifts from 0.1 to 0.2 for HI observations. Previous data show that these kinds of strong RFI will occupy nearly half of the observation time. Considering the RFI contamination and sensitivity limitation, CRAFTS will be difficult to detect galaxies at redshifts above 0.1. If we only consider the HI galaxies at redshifts below 0.1, in Monte-Carlo simulation, CRAFTS will be able to detect approximately $2.8\times10^5$ and $4.4\times10^5$ galaxies for one-pass and two-pass survey, with a mean redshift of 0.04 and 0.05, respectively.

\subsection{Compared with other planned large-scale HI surveys}

Apart from FAST, several large telescopes including ASKAP (the Australian SKA Pathfinder; \citealt{Johnston2007, Johnston2008}), MeerKAT (the South African Meer-Karoo Array Telescope; \citealt{Jonas2016}) and Apertif (APERture Tile In Focus) upgradation of WSRT (Westerbork Synthesis Radio Telescope; \citealt{Oosterloo2010}) also plan to complement large-scale HI surveys.

Those surveys can be divided into two types. One is to target a small area with long integration time to achieve depth. LADUMA\footnote{http://www.laduma.uct.ac.za/} (Looking at the Distant Universe with MeerKAT Array; \citealt{Blyth2016}) intends to study the cosmic evolution of HI galaxies to a redshift of $z$$\thicksim$1.4 with a single pointing of 3424 h; MIGHTEE-HI\footnote{http://idia.ac.za/mightee/}  (HI component of the "MeerKAT International GHz Tiered Extragalactic Exploration" survey; \citealt{Jarvis2016}) aims to cover a sky coverage of 20 $\mathrm{ deg^{2} }$ to $z$$\thicksim$0.5 with 16 h per pointing; Apertif\footnote{http://old.astron.nl/radio-observatory/apertif-surveys \label{fn:apertif}} plans to complement a medium-deep survey covering $\mathrm{\thicksim450 \, deg^{2}}$ to a redshift of 0.26, observed for 7$\times$12 h; carried by ASKAP, DINGO\footnote{https://dingo-survey.org/survey-design/} (Deep Investigation of Neutral Gas Origins; \citealt{Johnston2008, Duffy2012b}) is arranged in two tiers, named as DINGO-Deep and DINGO-Ultradeep, which plans to cover 150 $\mathrm{deg^{2}}$ (5$\times$500 h) to redshifts $0<z<0.26$ and 60 $\mathrm{deg^{2}}$ (2$\times$2500 h) to redshifts $0.1<z<0.43$, respectively. The other type is relatively shallow of large sky areas. WNSHS\footref{fn:apertif}\footnote{http://www.astron.nl/phiscc2014/Documents/Surveys/jozsa$\_$dwingeloo$\_$wnshs.pdf} (the Westerbork Northern Sky HI Survey; \citealt{Koribalski2012}) aims to cover $\thicksim$3500 $\mathrm{deg^{2}}$ to a redshift of 0.26 with 12 h per pointing, detecting about 90,000 galaxies out to a redshift of z$\thicksim$0.1; WALLABY\footnote{https://www.atnf.csiro.au/research/WALLABY/} (The Widefield ASKAP L-band Legacy All-sky Blind Survey; \citealt{Johnston2008, Duffy2012b,Koribalski2020}) plans to observe 75$\%$ of the sky (-90\degree<DEC<+30\degree) to redshifts of z$\lesssim$0.26, detecting $\thicksim$500,000 HI galaxies.  

As discussed previously, FAST will be easily contaminated by source confusion while observing HI galaxies at high redshifts compared to interferometers, limited by its relatively large beam size (e.g. the resolution of WALLABY is $\thicksim$30 arcsec). However, a major advantage of FAST is its large collecting area, which allows the detection of very low mass galaxies nearby. CRAFTS (-14\degree<DEC<+66\degree) and WALLABY (-90\degree<DEC<+30\degree) cover approximately $95\%$ of the sky in total, the combination of their final catalogues will enlarge the census of HI galaxies unprecedentedly, which is excellent in investigating the statistical properties of HI galaxies in the local universe. The combined observational data from CRAFTS and WALLABY will benefit from the higher resolution of ASKAP and better sensitivity of FAST, which will give a full sampling of nearby faint HI galaxies and their surroundings (e.g. \citealt{Koribalski2020}).

\section{Conclusions}
\label{sec:sec8}

In this paper, we have briefly introduced the technical parameters of CRAFTS extragalactic HI galaxy survey derived from commissioning observations, and predicted the potential of the survey on HI galaxy detection and HIMF measurements. We summarized our main results as the following:

\begin{enumerate}

\item CRAFTS will be able to detect approximately $2.9\times10^5$ and $4.8\times10^{5}$ galaxies ranging from $10^{6.4}$ to $10^{11}\,\mathrm{M_{\odot}}$ for one-pass and two-pass survey, with a mean redshift of nearly 0.047 and 0.055, respectively. Considering the RFI situation and sensitivity limitation, CRAFTS will effectively survey HI galaxies at redshifts below 0.1. \\

\item CRAFTS will recover the "faint end" slope of nearby galaxies to $\mathrm{10^{7}\,M_{\odot}} $ and measure the global "knee mass" to a redshift of 0.1.\\

\item From our simulation, source confusion would reduce approximately $2\%$ of detections for CRAFTS. After considering source confusion effects, the measured "knee mass" of the HIMF would be overestimated by 1$\sigma$ and 4$\sigma$ Poisson counting error at redshifts of 0$-$0.05 and 0.05$-$0.10, respectively. \\

\item The large survey area and outstanding sensitivity performance enable CRAFTS to better explore the galaxy interactions in different environments and the spatial distribution of HI galaxies in the local universe. CRAFTS will also enlarge the census of low-mass galaxies and constrain the "faint end" slope of nearby HIMF,  shedding some light on the "missing satellite" problem.

\end{enumerate}

\section*{Acknowledgements}

The author would like to thank helpful discussion with Marko Kc\v{r}o, Michael G. Jones, Ying Zu and Sean E. Lake. This work is supported by the National Natural Science Foundation of China grant No. 11988101, No. 11725313, No. 11673029, No. 11590783, No. 11973051; the National Key R$\&$D Program of China grant No. 2017YFA0402600, No. 2016YFA0400702; the Open Project Program of the Key Laboratory of FAST, NAOC, Chinese Academy of Sciences; the International Partnership Program of Chinese Academy of Sciences grant No.  114A11KYSB20160008. JW thank the support from National Natural Science Foundation of China grant No. 12073002 and No. 11721303. JF acknowledges the support by the Youth innovation Promotion Association CAS and Shanghai Committee of Science and Technology grant No. 19ZR1466700. 
%%%%%%%%%%%%%%%%%%%%%%%%%%%%%%%%%%%%%%%%%%%%%%%%%%

\section*{Data Availability}

The data underlying this article will be shared on reasonable requests to the corresponding authors.

%%%%%%%%%%%%%%%%%%%%%%%%%%%%%%%%%%%%%%%%%%%%%%%%%%

\bsp	
\label{lastpage}
\end{document}